\documentclass[twocolumn,aps,prstab,groupedaddress,longbibliography]{revtex4-1}
\pdfoutput=1 

\usepackage{gensymb}
\usepackage{mathptmx}
\usepackage{graphicx}
\usepackage{caption}
\usepackage{subcaption}

\usepackage{threeparttable} 
\usepackage{textcomp}
\begin{document}

\title{Design and operation of the air-cooled beam dump for the extraction line of CERN's Proton Synchrotron Booster}


\author{A. Perillo-Marcone}
\email[]{antonio.perillo-marcone@cern.ch}
\affiliation{CERN, 1211 Geneva 23, Switzerland}
\author{M. Calviani}
\author{N. Solieri}
\author{A. Ciccotelli}
\author{P. Kaiser}
\author{A. Sarrio}
\author{V. Venturi}
\author{V. Vlachoudis}



\begin{abstract}
A new beam dump has been designed, built, installed and operated to withstand the future proton beam extracted from the Proton Synchrotron Booster (PSB) in the framework of the LHC Injector Upgrade (LIU) Project at CERN. The future proton beam consists of up to 1$\times$10$^{14}$ protons per pulse at 2 GeV and is foreseen after the machine upgrades planned for CERN's Long Shutdown 2 (2019-2020).

In order to be able to efficiently dissipate the heat deposited by the primary beam, the new dump was designed as a cylindrical block assembly, made out of a copper alloy and cooled by forced airflow.

In order to determine the energy density distribution deposited by the beam in the dump, Monte Carlo simulations were performed using the FLUKA code, and thermo-mechanical analyses were carried out by importing the energy density into ANSYS\textsuperscript{\textregistered}. In addition, Computational Fluid Dynamics (CFD) simulations of the airflow were performed in order to accurately estimate the heat transfer convection coefficient on the surface of the dump.

This paper describes the design process, highlights the constraints and challenges of integrating a new dump for increased beam power into the existing facility and provides data on the operation of the dump.
\end{abstract}





\maketitle
\flushbottom

\section{Introduction}
The Proton Synchrotron Booster (PSB) has accelerated protons as part of CERN's accelerator complex for more than 40 years~\cite{baribaud_800_1973}. Ever since its construction, the accelerator has undergone several upgrades that have made possible, among other things, drastic increases in both the extracted intensities and extraction energies. From the 5.4$\times$10$^{12}$ protons per pulse (ppp) extracted at 800~MeV in late 1973~\cite{hanke_past_2013}, in recent years the Booster has accelerated up to 4$\times$10$^{13}$ ppp at an energy of 1.4~GeV. Moreover, as a result of the LHC Injector Upgrade (LIU) project at CERN~\cite{Damerau:1976692}, foreseen after the machine upgrades planned for CERN's LS2 (Long Shutdown 2, 2019-2020), the Booster will be able to accelerate up to 1$\times$10$^{14}$ ppp at 2~GeV.

It was in the framework of this series of upgrades that the dump was replaced in October 2013, during CERN's LS1 (Long Shutdown 1, 2013-2014). The previous dump had in fact been operating since the construction of the PSB in 1972 and could no longer safely absorb the beams of increased intensity and energy accelerated by the booster.

The aim of this work is to present the R\&D activities for the new design of the dump, describing how it can cope with the upgraded beam parameters, as well as the procedure for its replacement, taking into consideration the radiological requirements along with the physical and infrastructural constraints inherent of the project. The operational feedback from the use of the dump between LS1 and 
LS2 (2015 - 2018) is also compared with the finite-element simulations that guided the design of the new dump.

\subsection{Original dump core} \label{previous-dump-core}
\begin{figure}[ht]
\centering
\includegraphics[width=0.45\textwidth]{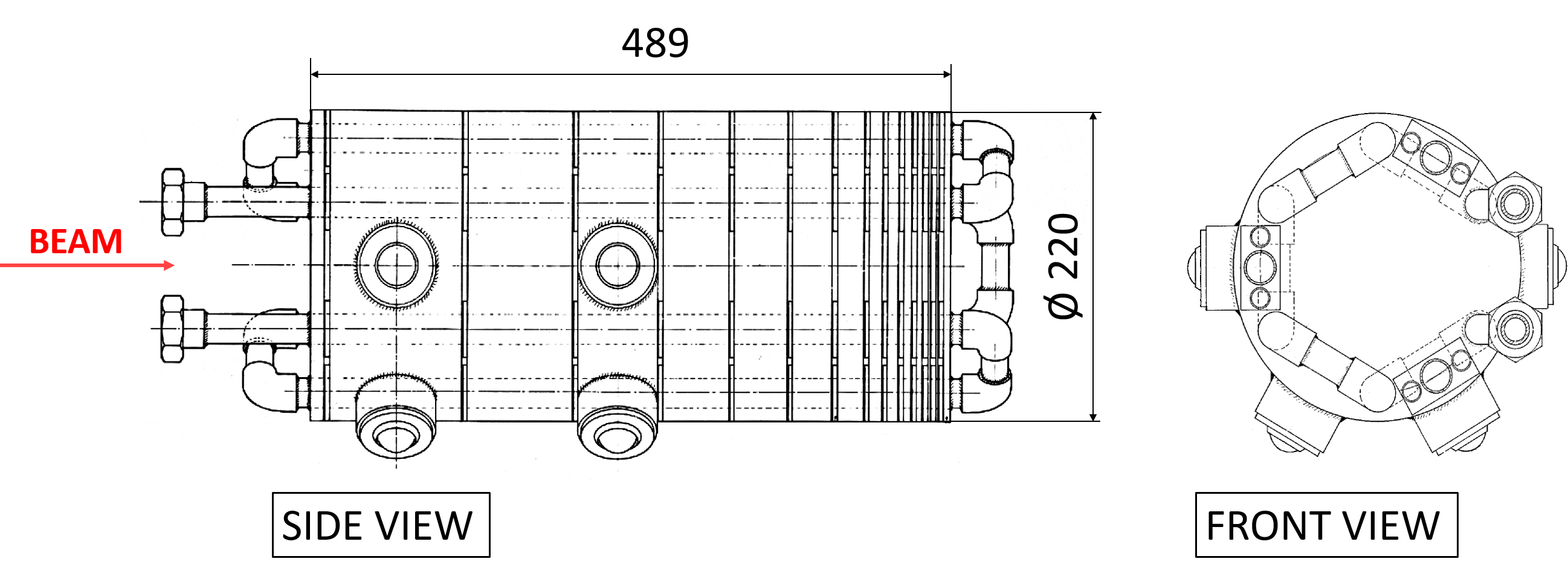}
\caption{Drawing of the original PS Booster dump core, showing the side and front view of the assembly. Dimensions are reported in mm. }
\label{fig:old-dump-drawing}
\end{figure}

As shown in Figure~\ref{fig:old-dump-drawing}, the PS Booster original dump core consisted of a series of 13 Fe37-steel~\cite{ISO9930} disks, assembled in decreasing order of thickness along the beam axis, from 100 down to 2~mm, with a constant gap of 4~mm between each of them~\cite{sarrio_martinez_psb_nodate}. This assembly had a total length of 489~mm and a diameter of 220~mm.

The dump was water-cooled by a single  stainless steel cooling pipe running forwards and backwards six times through the disks at different angular positions. 
In the last years of operation, however, the cooling pipes were disconnected when water leaks were detected. Since the dump core was not under vacuum and was exposed to the atmosphere, limited natural air convection was left as the only means of evacuating the heat deposited in the core. Moreover, the beam pipe that was inserted in the cavity leading up to the dump core experienced vacuum leaks. As a result of this, it was detached from the beam line and then disconnected from the vacuum system~\cite{sarrio_martinez_psb_nodate}.

The obsolescence of the dump as well as the limitations induced by the reduced cooling performance after these events forced the design, construction and installation of a new generation beam absorber, capable of coping with the requirements of the LIU Project.

\subsection{Dump Area Layout}
\label{dump-area-layout}
\begin{figure}[ht]
\centering
\includegraphics[width=0.4\textwidth]{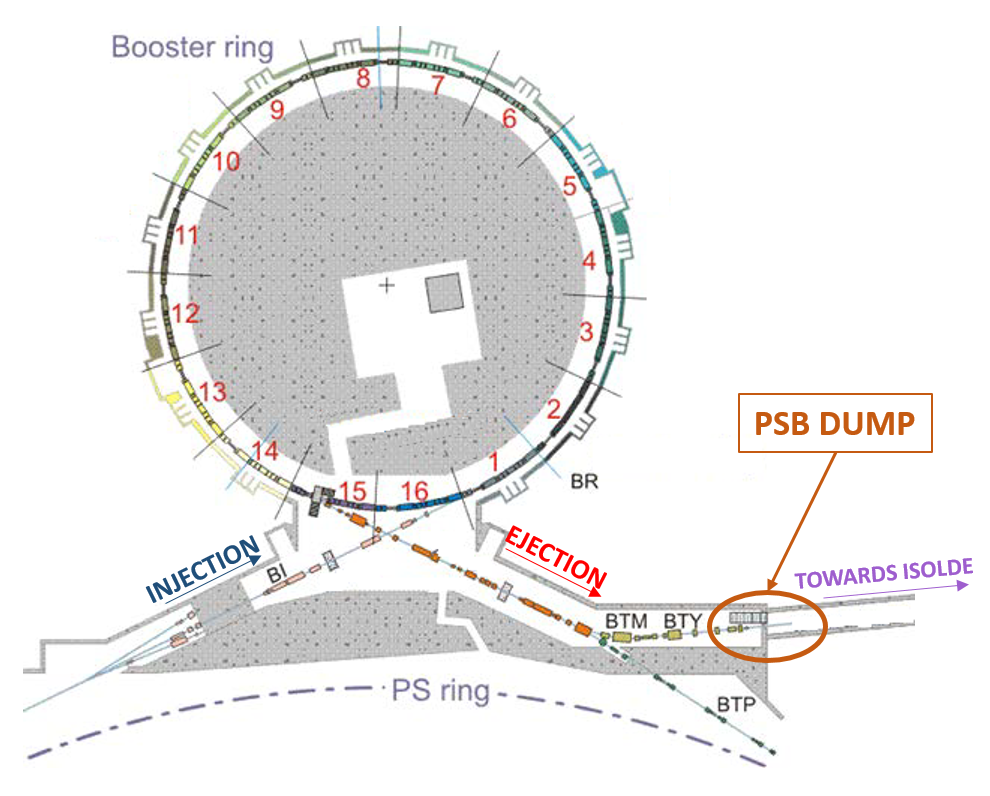}
\caption{Layout of the Proton Synchrotron Booster (PSB). The PSB dump is located at the end of the BTM line, below the transfer line to the ISOLDE Facility and after the ejection line towards the Proton Synchrotron (PS).}
\label{fig:PSB Layout}
\end{figure}

\begin{figure}[ht]
\centering
\includegraphics[width=0.45\textwidth]{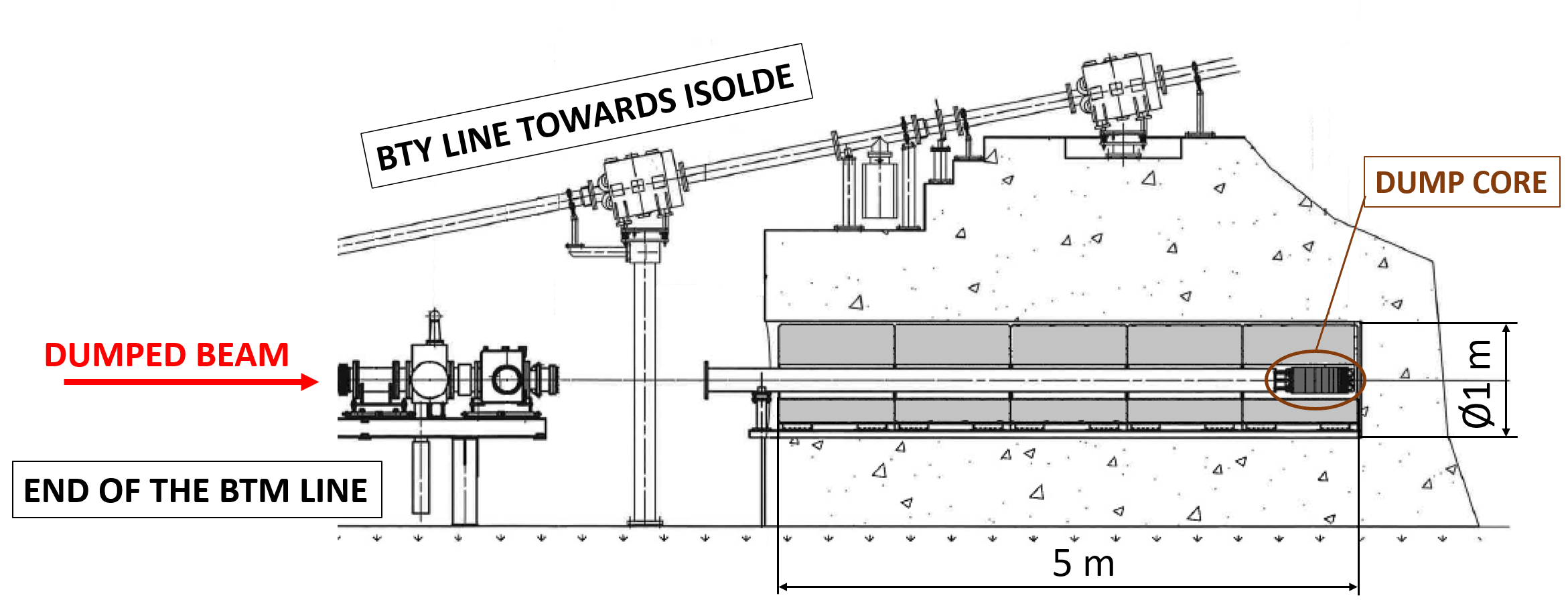}
\caption{Layout of the PSB Dump area. On the left of the picture, the end of the BTM line and the BTY line towards ISOLDE. On the right, the original dump core, its shielding and beam pipe inserted in the dump cavity.}
\label{fig:location}
\end{figure}

The PSB external dump is located at the end of the BTM line, after the PSB-to-PS extraction line and below the transfer line feeding the ISOLDE facility (see Figure~\ref{fig:PSB Layout}). As can be seen in Figure~\ref{fig:location}, the original dump core was installed inside a 5~m-deep, 1~m-diameter cavity shielded by hollow-cylinder shaped concrete blocks.

Due to limitations on the possible interventions in the PSB extraction area, the new dump core had to be installed in the same cavity as the original one~\cite{sarrio_martinez_psb_nodate}. The compatibility with the installation inside the cavity was, therefore, one of the main factors that drove the design phase for the new dump core and its shielding.

\section{Beam Parameters}
\label{sec:beampar}
Based on past experience~\cite{bartmann_ps_2012}, around 6\% of the beams extracted from the PSB are regularly dumped during normal operation. Considering conservatively 10\% of the extracted protons to be dumped, as well as the increase of the intensity due to the installation of the Linac4~\cite{Lombardi2019}, around 5$\times$10$^{19}$ protons will reach the dump core each year.

Moreover, in the case of commissioning periods, up to 50\% of the beams that are extracted from the PSB are sent to the dump for several consecutive months. This is the case, for example, of the commissioning period that is foreseen after the connection of the Linac4 to the PSB and the consequent upgrade to 2 GeV beam energy.

Out of all the possible types of beams extracted from the PSB after LS2, two were identified as the most critical for the operation of the dump: the NORMGPS and the LHC25ns. The NORMGPS is a standard beam produced for the operation of the GPS (i.e. General Purpose Separator), which is one of the two isotope separators of the ISOLDE Facility at CERN (the other one being the High Resolution Separator, HRS). The LHC25ns is instead the standard beam that is employed for feeding the Large Hadron Collider (LHC). Their main parameters are listed in Table~\ref{tab:design-parameter}.
As it can be seen, the NORMGPS beam has a higher intensity and deposits, therefore, a higher total energy in the dump. The LHC25ns, on the other hand, despite having a lower intensity,  produces a higher maximum energy deposition density into the dump core due to its smaller transverse size.
The new beam dump was designed to last for the residual facility lifetime of 25 to 30 years while operating under the conditions produced by these beams, excluding any further upgrade of the accelerators. During operation, the characteristics of the beam that are extracted to the dump are monitored using beam instrumentation that is placed in close proximity of the dump cavity. In particular, the last Beam Position Monitor (BPM) and Beam Current Transformer (BCT) are placed, respectively, less than 4 and 2~m away from the entrance of the dump cavity. The profile of the beam is also monitored by means of 3 Secondary EMission (SEM) grids.

\begin{table*}[htbp]
\centering
\caption{Parameters of the most critical beams for the operation of the dump, after LS2: NORMGPS and LHC25ns~\cite{bartmann_ps_2012}.}
\label{tab:design-parameter}
\begin{tabular}{lrr}
\hline\hline
                              & NORMGPS     & LHC25ns     \\\hline
Max beam intensity [ppp]           & $1\times10^{14}$        & $1.4\times10^{13}$      \\
Energy    [GeV/pp]               & \multicolumn{2}{c}{\hspace{15pt} 2 GeV} \\
Pulse Length     [ns]             & 940       & 1701      \\
Number of bunches             & 4           & 4           \\
Full bunch length [ns]         & 160       & 180       \\
Spacing between bunches [ns] & 100       & 327       \\
Intensity per bunch  [p]      & $2.5\times10^{13}$      & $3.25\times10^{12}$     \\
Beam size ($1\sigma, H\times V$) [mm] & $13\times13$       & $5.1\times5.1$     \\
Brightness                    & 0.5         & 3.7        \\\hline
\end{tabular}%
\end{table*}

\section{Design of the new dump}
\subsection{Dump core design}
\label{sec:dump-core-design}

As mentioned in Sec.~\ref{dump-area-layout}, the only possible location for the installation of the new dump was the same dump cavity at the end of the BTM line where the original dump was installed. Given these physical constraints, the structure of the new core-shielding assembly had to be similar to the original one. It consists of a cylindrical dump core that is installed at the downstream end of the dump cavity and surrounded by a set of cylindrical shielding blocks. The materials, cooling technology and dimensions of the new core are, however, very different from the first generation one.

As described in Sec.~\ref{previous-dump-core}, the previous core was made up of a series of steel disks, with a total length of 490~mm. Due to the higher energy of the upgraded beam, however, a longer dump made of a higher density material was required in order to contain most of the prompt radiation produced in the beam interaction process and reduce leaking of radiation downstream. Copper alloys, thanks to their high density and high thermal conductivity, were ideal candidate materials. Considering a dump entirely made of copper, the nuclear inelastic scattering length ($\lambda$) of 2~GeV protons is 13.8~cm. A dump length of at least 140~cm, corresponding to roughly 10$\lambda$, was therefore required to fully contain the impacting beam and reduce the uncollided proton fraction escaping the dump down to less than 5x10$^{-5}$.

Since the temperatures and stresses resulting from the energy deposited in the dump by the beam would have been too high for a long term reliable operation of pure copper (Cu-OFE), a Copper-Chromium-Zirconium (CuCr1Zr - UNS C18150 \cite{deutsches_kupferinstitut_kupferdatenblatt_2005}) alloy was selected. Although the thermal conductivity of this alloy is slightly lower than that of pure copper, it features higher strength at the expected operating temperatures, allowing for both peak temperatures and stresses to be safely maintained within the limits of this material. This absorbing material has been chosen also for other future facilities such as the European Spallation Source (ESS) tuning dump~\cite{Garoby_2017,Lee2017}.

The diameter of the dump core was required~\cite{bartmann_ps_2012} to contain up to 5$\sigma$ of the maximum beam size. According to the estimations of beam size variability reported in~\cite{bartmann_ps_2012},  the maximum beam size corresponds to the 1.0 GeV NORMGPS/NORMHRS beam, calculated to measure 185 and 192 mm at 5$\sigma$ in the horizontal and vertical plane, respectively.    



\begin{figure}[ht]
\centering
\includegraphics[width=0.45\textwidth]{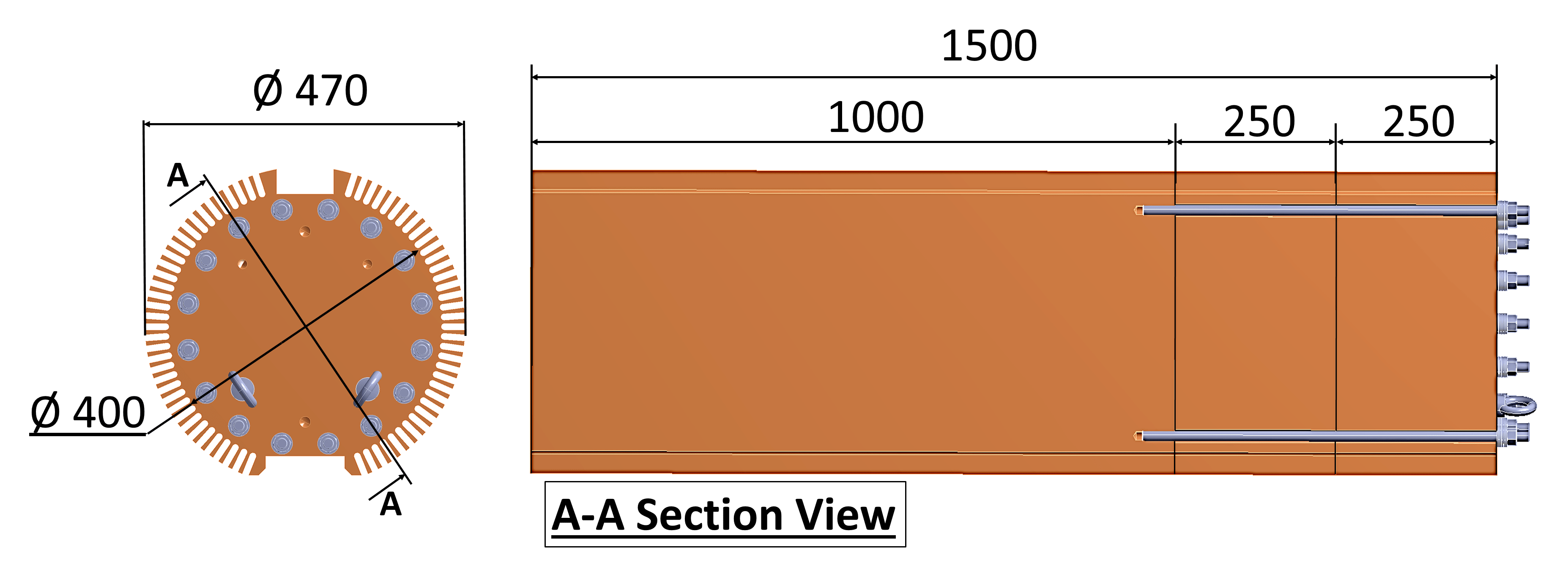}
\caption{(left) Front view and (right) section view of the CAD model of the new CuCr1Zr PSB dump core. Dimensions are reported in mm.}
\label{fig:dump-core}
\end{figure}

A CAD model of the new dump core design that resulted from taking into account these requirements is shown in Figure~\ref{fig:dump-core}. As can be seen, the length of the core was increased from 490 to 1500~mm, while its diameter was also increased from 220 to 400~mm. The lateral surface of the core features a dense series of fins, which, as it will be further detailed in Sec.~\ref{sec:cooling}, contribute to increase the heat transfer surface between the dump core and its air cooling system by a factor 5.

Due to size constraints imposed by the Cu-alloy manufacturer, the dump core was split longitudinally into three cylindrical blocks of equal diameter. These parts were then clamped together by means of threaded rods and spring washers in order to ensure good thermal contact. Two lifting rings were screwed onto the front face of the dump core. Stainless steel cables were attached to these rings so that the dump could be easily extracted in the future.

\subsection{Dump core-shielding assembly}
\begin{figure}[ht]
\centering
\includegraphics[width=0.45\textwidth]{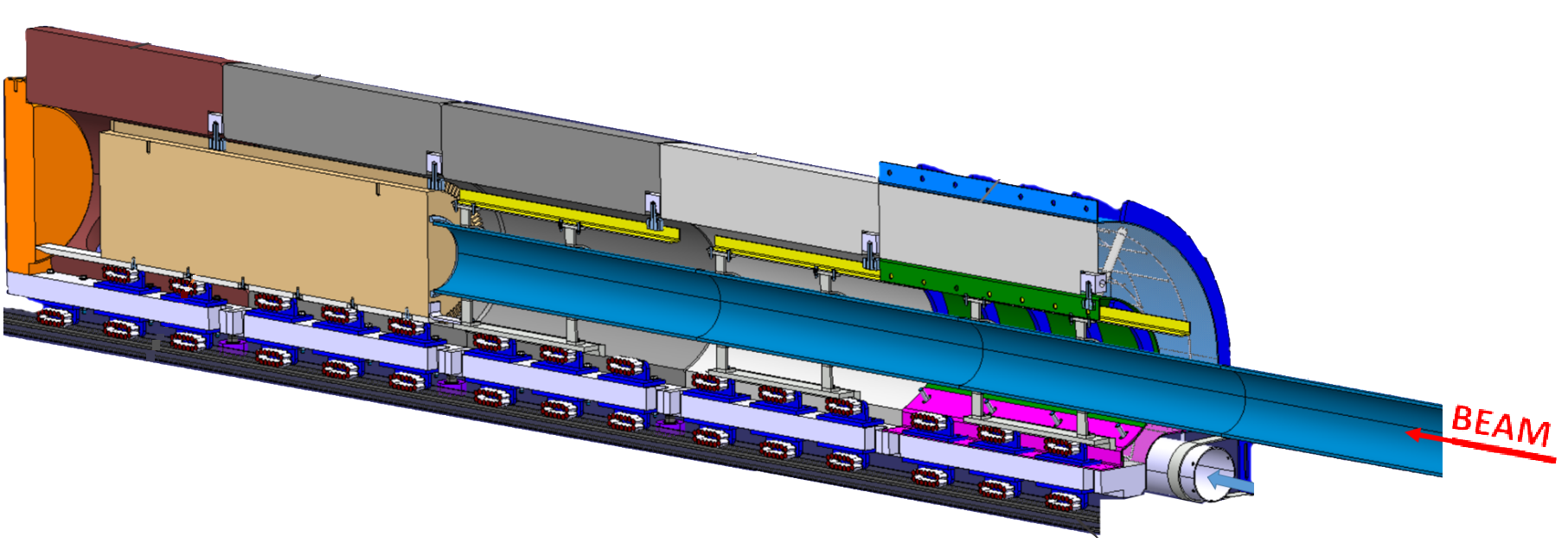}
\caption{CAD Section view of the new dump core-shielding assembly, showing the CuCr1Zr dump core, the cast iron shielding blocks as well as the concrete ones. The ducts for the air cooling and the aluminum alloy beam pipe are also visible.}
\label{fig:dump-assembly}
\end{figure}

The current dump core-shielding assembly is shown in a CAD section view in Figure~\ref{fig:dump-assembly}. As described in Sec.~\ref{sec:dump-core-design}, this assembly was designed to be installed in the same dump cavity that has been used since the installation of the PSB.

Since the new dump core diameter had to be larger than that of the previous one, the available space  for shielding blocks was smaller. A higher density material was therefore required to guarantee the same shielding effectiveness. Cast iron was used for the three downstream shielding blocks, while concrete was chosen for the two upstream ones (closer to the cavity entrance) in order to minimize the residual dose in the area outside of the cavity and to more effectively absorb the thermal neutrons backscattered from the absorber. As can be seen in Figure~\ref{fig:dump-assembly}, the five blocks have an annular geometry and cover the whole depth of the cavity. In the radial direction, a gap of 30~mm is left between the inner diameter of the blocks and the tip of the dump core's fins, so as to limit the value of the pressure drop experienced by the air flowing in the gap.

\begin{figure}[ht]
\centering
\includegraphics[width=0.35\textwidth]{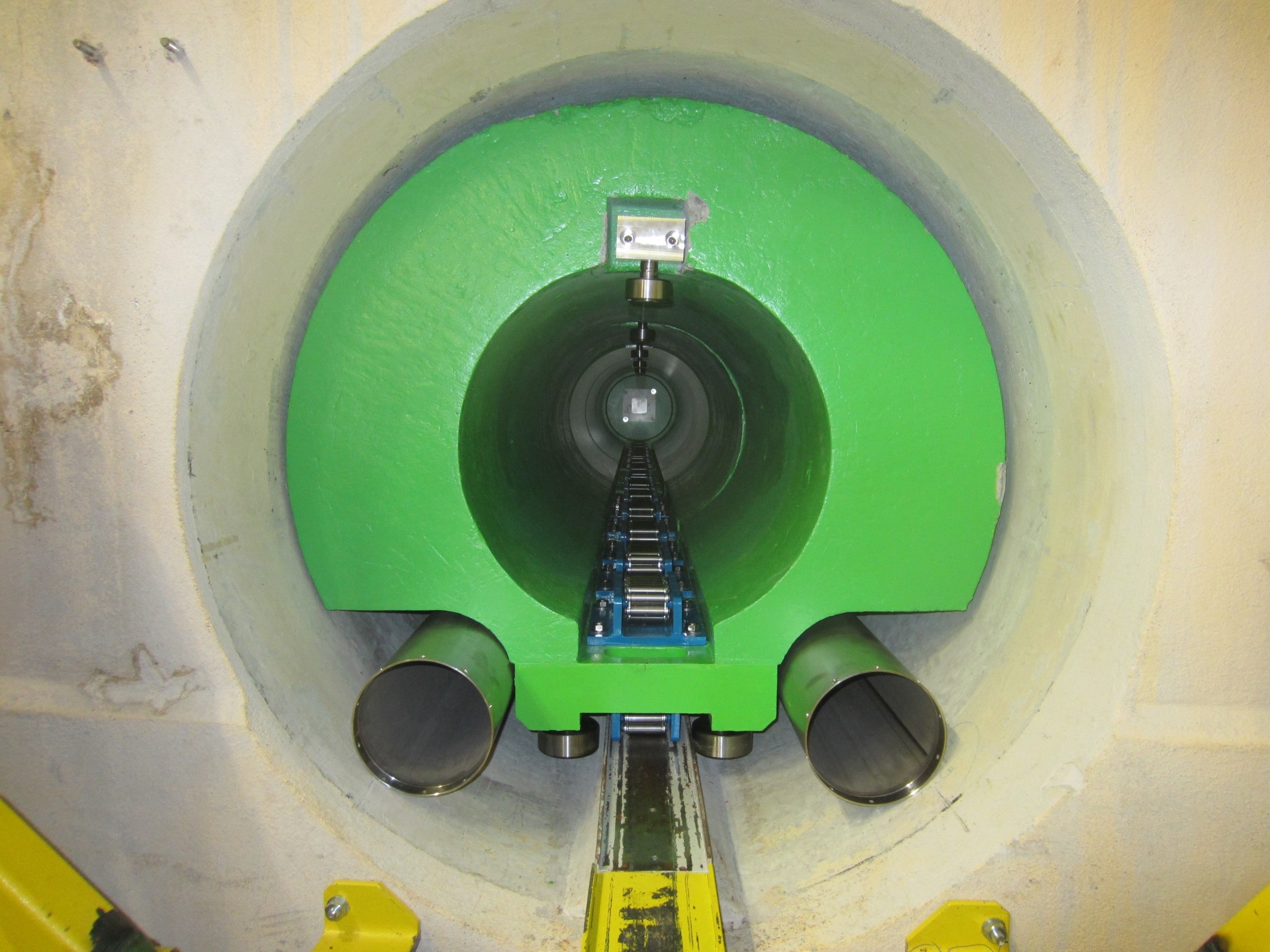}
\caption{Front photo of the PSB dump cavity, taken during the insertion of the new shielding blocks in October 2013. The concrete shielding blocks are visible, together with the two air inlet ducts as well as the lower and upper skates.}
\label{fig:shielding-assembly}
\end{figure}

As shown in Figure~\ref{fig:shielding-assembly}, each shielding block features two series of skates. The lower skates allow the blocks to slide onto a pre-existing steel rail, which is fixed onto the external concrete shielding of the dump cavity. The upper skates allow the dump core to reach its position at the downstream end of the cavity by sliding on the shielding blocks, once these have been inserted.

With the current configuration of the beamline, beams sent to the dump travel in air for approximately five metres. In order to minimize the radiological impact of this design, the air activated by the beam is confined in an aluminum alloy air pipe over almost this total length.

\subsection{Dump Cooling}
\label{sec:cooling}
Under the conditions mentioned in Sec.~\ref{sec:beampar}, the total average power that will be deposited by the beam in the dump core-shielding assembly will amount to 11~kW. As it is further detailed in Sec.~\ref{sec:en-dep}, this value was evaluated by means of simulations performed using the FLUKA~\cite{ferrari_fluka:_2005} Monte Carlo code and it corresponds to the sum of the power that is deposited in the dump core (9.7~kW) and in the shielding assembly (1.3~kW).

In order to dissipate this amount of power, either water or air would have been viable solutions for the cooling system. Due to the radiation protection challenges associated with the use of water, such as the higher production (and retention) of tritium and the danger posed by water leaks, air was chosen as the coolant for this application.

\begin{figure}[ht]
\centering
\includegraphics[width=0.45\textwidth]{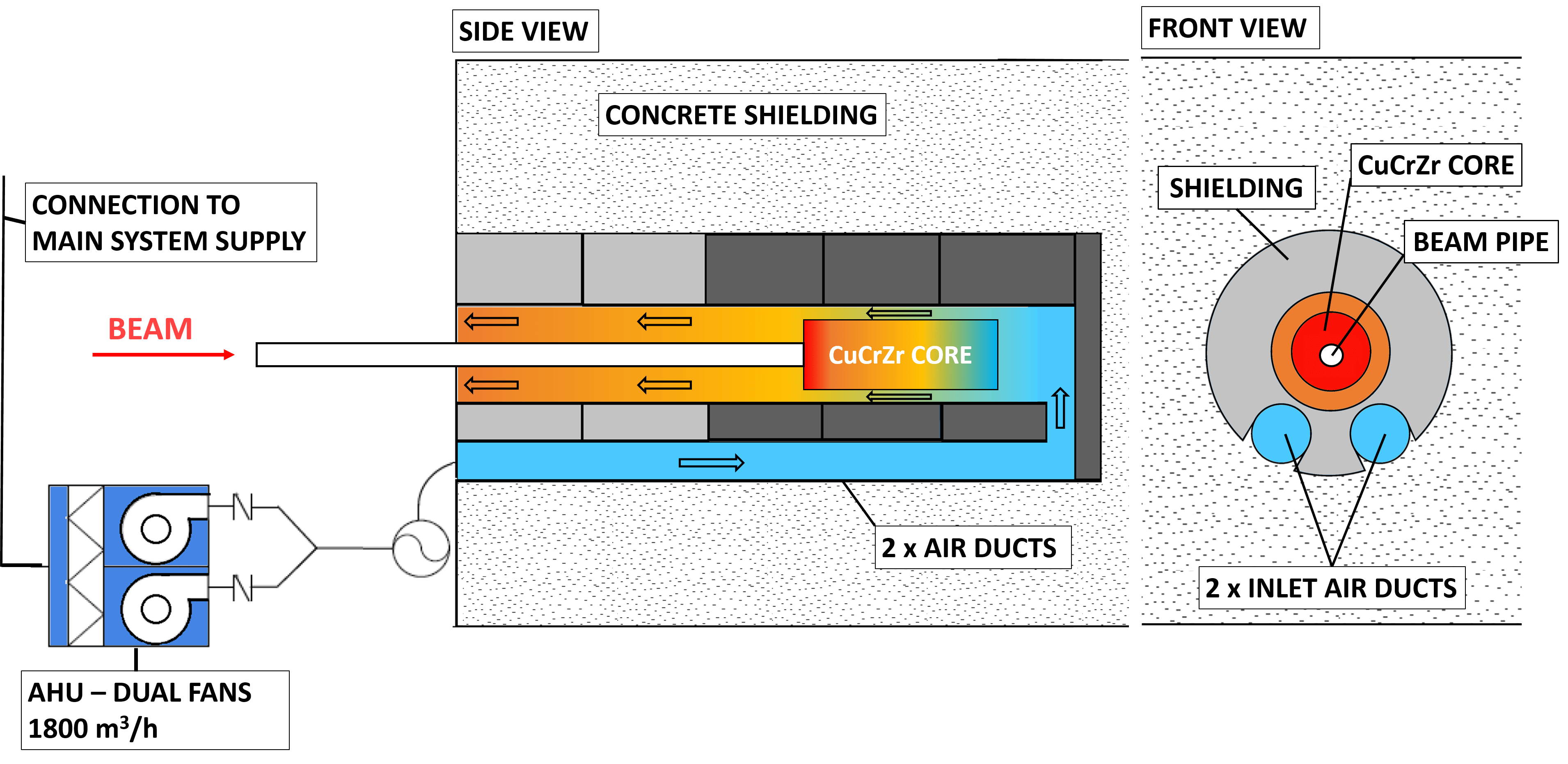}
\caption{Simplified plan of the cooling system of the new PSB dump, showing the different components that are part of the the system.}
\label{fig:design-cooling}
\end{figure}

Similarly to the original dump, it was not required for the new dump core to operate in vacuum. This greatly simplified the design of the cooling system (schematics shown in Figure~\ref{fig:design-cooling}). As can be seen, air is blown from the downstream end of the dump through two inlet ducts that are housed in the lower part of the shielding blocks (visible in Figure~\ref{fig:shielding-assembly}). Once it reaches the end of the cavity, the air is then forced to flow backwards between the fins placed along the lateral surface of the dump. Finally, the air flows out of the cavity into the tunnel, where it is removed by the existing tunnel ventilation~\cite{Mason2012}.

\begin{figure}[ht]
\centering
\includegraphics[width=0.4\textwidth]{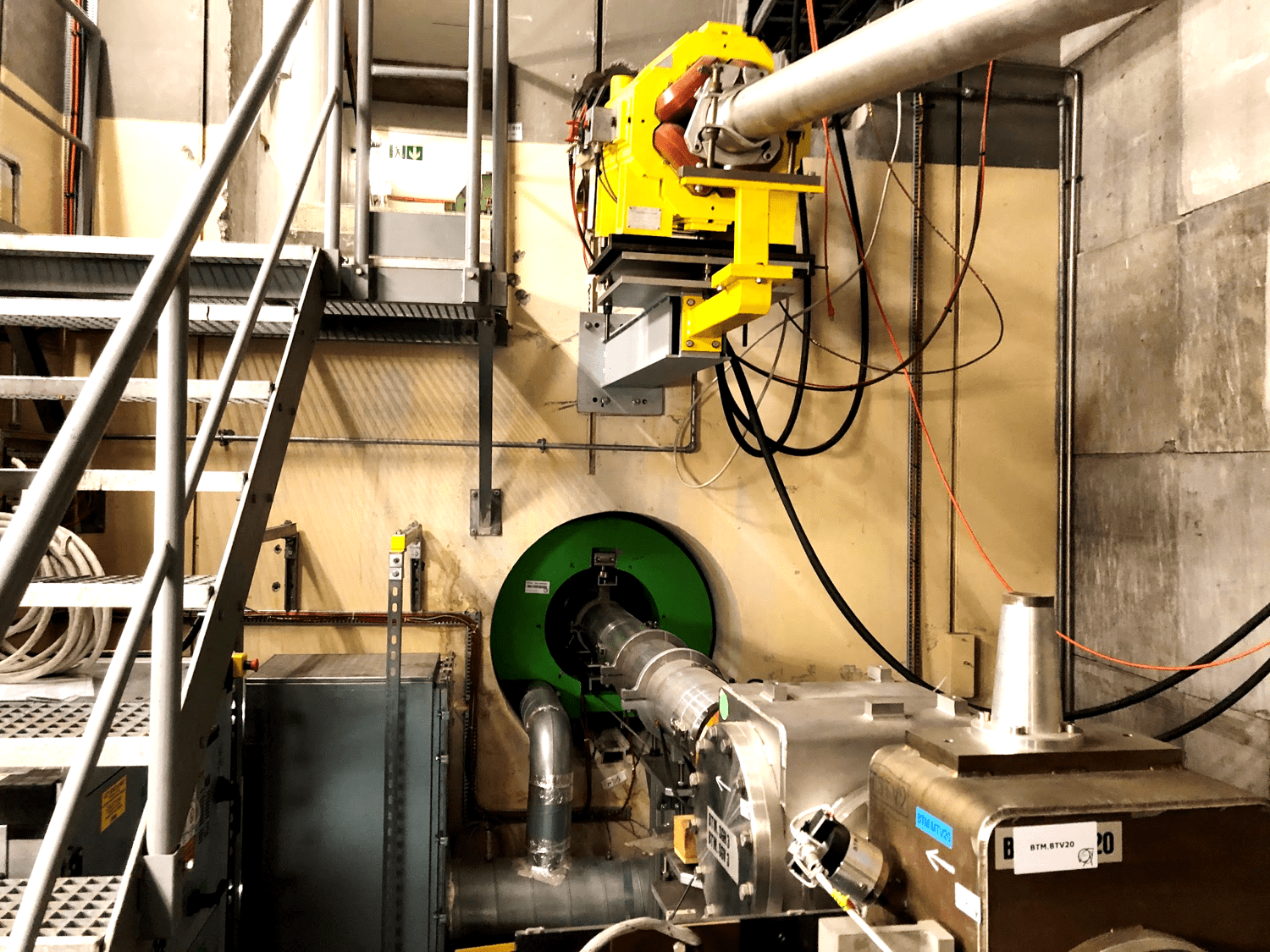}
\caption{View of the entrance of the beam dump cavity at the end of the BTM cavity in November 2019. Visible on the left side in the foreground under the staircase to ISOLDE, the Air Handling Unit (AHU) and the cooling ducts feeding into the cavity.}
\label{fig:cooling-set-up}
\end{figure}

As can be seen in Figure~\ref{fig:cooling-set-up}, the air handling unit (AHU) is located outside of the cavity in an area with low prompt radiation conditions, beside the beam line. It is composed of two independent fans in order to provide redundant operation in case of failure of one unit~\cite{Mason2012}.

The air supply is taken from the general ventilation system of the area, which has a theoretical maximum temperature of 20~\degree C. In order to maintain the dump core within acceptable temperatures and to keep the air temperature increase below 20~\degree C. The required flow rate is 1800~Nm$^3$/h (normal cubic meters per hour).

\section{Numerical Analyses}\label{FEM}
Analyses performed by means of finite-element simulations were fundamental in guiding the design of the dump core. These analyses were structured in the following steps: 

\begin{enumerate}
    \setlength{\itemsep}{0pt}%
    \setlength{\parskip}{0pt}%
    \item Evaluation of the energy deposited in the dump core by the proton beams by means of FLUKA~\cite{ferrari_fluka:_2005} Monte Carlo simulations;
    \item Interpolation of the FLUKA-calculated energy deposition in the dump core in an ANSYS-CFX\textsuperscript{\textregistered} CFD (i.e. Computational Fluid Dynamics) model to evaluate the performance of the cooling system;
    \item Extraction of the HTC (i.e. Heat Transfer Coefficient) between air and dump core from ANSYS-CFX\textsuperscript{\textregistered} and its interpolation into an ANSYS-Mechanical\textsuperscript{\textregistered} model to assess the thermal-mechanical response of the dump core.
\end{enumerate}

This process was followed for each of the analyzed beams,  intensities and dumping rates.

\subsection{Energy deposition}
\label{sec:en-dep}
As previously introduced, the energy deposited by the beams into the dump core was computed by means of FLUKA~\cite{ferrari_fluka:_2005} Monte Carlo simulations. The characteristics of each beam were set consistently with those listed in Table \ref{tab:design-parameter}, which were derived from the study in \cite{bartmann_ps_2012}.

\begin{figure*}[ht]
\centering
\includegraphics[width=0.85\textwidth]{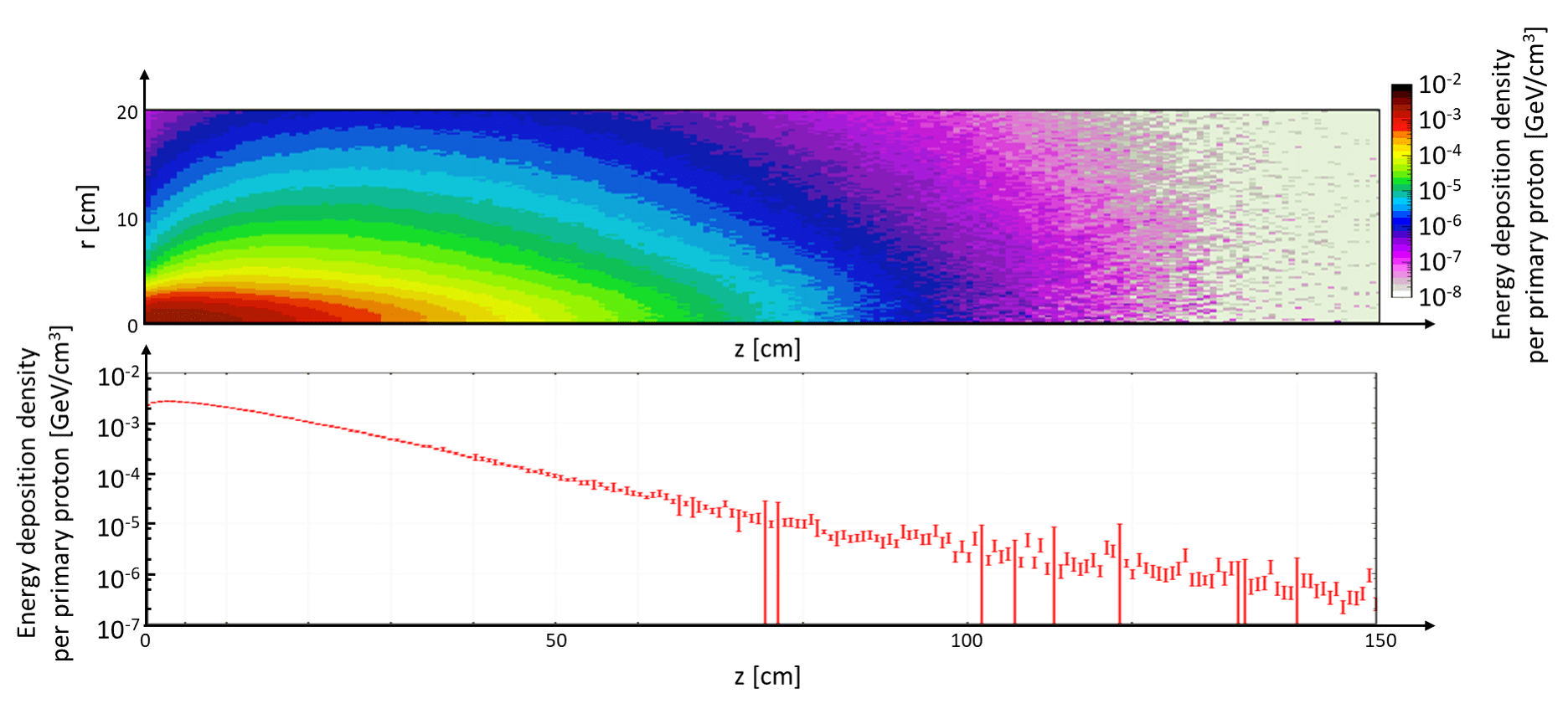}
\caption{Top: energy deposition density (GeV/$cm^3$ per primary proton) deposited into the dump core by the 2 GeV NORMGPS beam. Bottom: corresponding maximum value of the energy deposition density along the length of the dump core.}
\label{fig:energy-deposition}
\end{figure*}

Figure~\ref{fig:energy-deposition} shows the distribution of energy density deposited into the dump core by the 2~GeV NORMGPS beam. As  can be seen, a high energy deposition density is present in close proximity to the upstream face of the dump. This was the main reason why a material with high thermal conductivity, such as CuCr1Zr, was chosen to efficiently evacuate the heat towards the outer air-cooled surfaces of the core.

\subsection{CFD Simulations}
\label{sec:CFD}
Once the FLUKA-computed energy deposition into the dump core was obtained, it was interpolated for each case into an ANSYS-CFX\textsuperscript{\textregistered} steady state CFD model. This process was first performed in order to guide the design of the dump core and, in particular, to identify the optimal geometry of the fins on its lateral surface. As described in \cite{mason_liu-psb_2013}, the final fin geometry was chosen as a compromise between the maximization of the heat transfer surface and the minimization of the pressure drop. A smaller gap between consecutive fins would have allowed more fins to be placed on the lateral surface of the dump core, thereby increasing the heat transfer surface. This would have also, however, increased the value of the pressure drop experienced by the air flowing in the narrower gaps beyond what a AHU could handle for the specified flow rate of 1800~Nm\textsuperscript{3}/h.
The CFD-optimized dump core geometry features fins with a height of 35~mm, a gap between two consecutive fins at the level of their base of 10~mm and a thickness varying between 4 and 6.5~mm.

\begin{figure}[ht]
\centering
\includegraphics[width=0.45\textwidth]{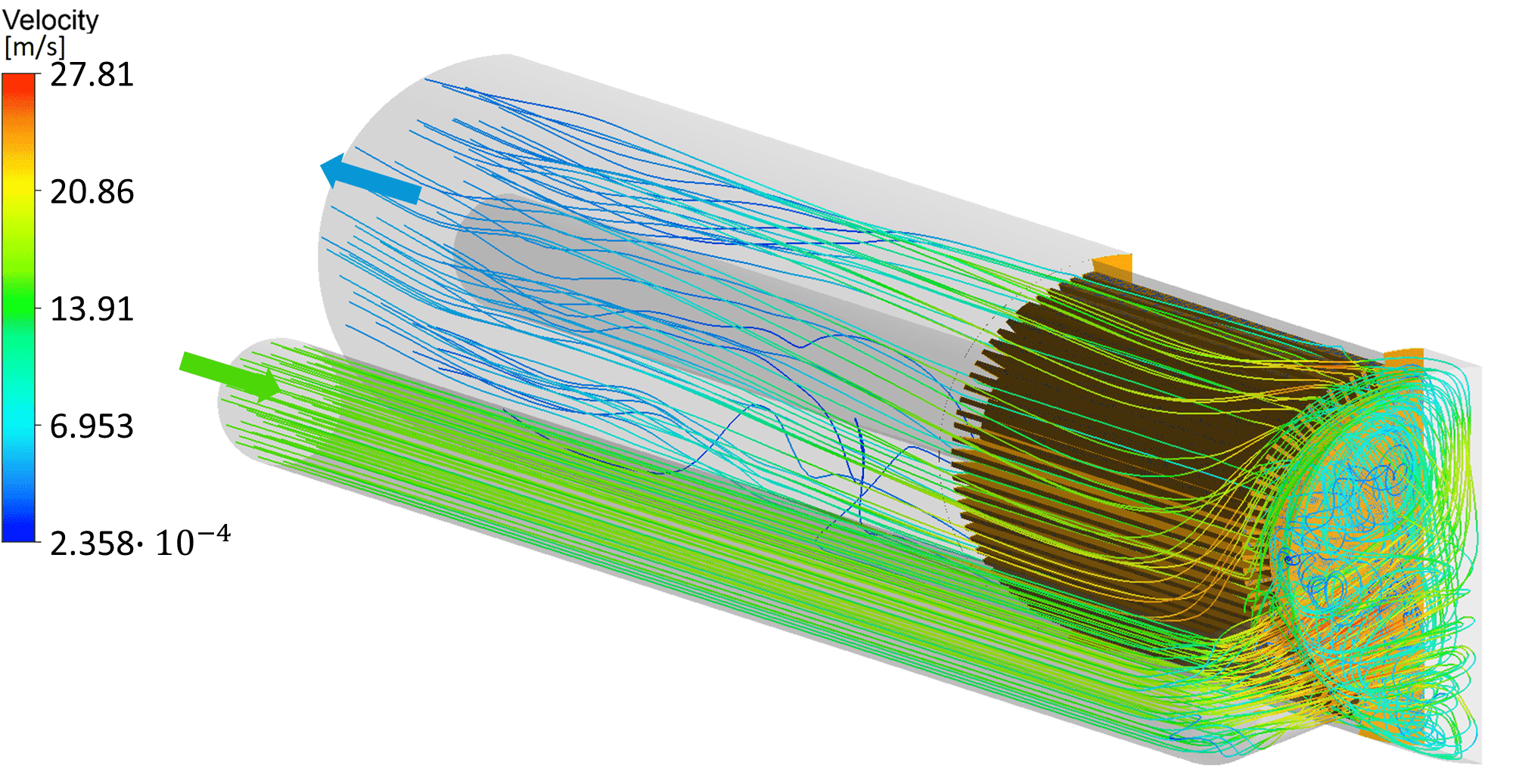}
\caption{Velocity streamlines obtained for the steady state CFD simulation that computes the impact of the 2~GeV NORMGPS beam with an intensity of $1\times 10\textsuperscript{14}$ ppp and a 50\% dumping rate (the nominal operation for which the dump was designed).}
\label{fig:CFX}
\end{figure}

The CFD model for the final geometry of the dump core is shown in Figure~\ref{fig:CFX}. As  can be seen, a symmetry boundary condition is implemented in the model due to the fact that the geometry, the thermal load induced by the beam impact and the air cooling are all nominally symmetric with respect to the central vertical plane. The heat transfer between air and shielding, as well as the energy deposited inside of the shielding were neglected for the purpose of this simulation. A Shear Stress Transport (SST) turbulence model was used to accurately resolve the thermal boundary layer in proximity of the surfaces of the core, for an accurate evaluation of the conjugate heat transfer between the CuCr1Zr core and the air.
In the case of  a 2~GeV NORMGPS beam with an intensity of 1$\times$10$\textsuperscript{14}$~ppp and a 50\% dumping rate (the nominal operation for which the dump was designed) the pressure drop  in the entire cooling system is 500 Pa and the increase in air temperature is 13 \degree C. 


\subsection{Thermal response of the dump core to high-intensity 2 GeV beams}
\label{sec:Thermal-analysis}
The FEM model was then applied to evaluate the thermal response of the dump core in the conditions that would determine the highest temperatures in currently-foreseen operation. These conditions are achieved during the previously mentioned prolonged dumping of a 2~GeV NORMGPS beam with an intensity of 1$\times$10$\textsuperscript{14}$~ppp at a 50\% dumping rate. Each beam pulse deposits 23.2~kJ in the dump core. Considering the repetition rate of 1.2~s and the 50\% dumping rate, this corresponds to an average deposited power of 9.67~kW. 
As specified in Sec.~\ref{sec:dump-core-design}, the dump core was longitudinally split into three cylindrical blocks, which were clamped together by means of threaded rods. This method of assembling causes some uncertainties in modelling the thermal contact between these blocks. It was therefore interesting to examine the response of the dump core in the two extreme cases between which the real response of the blocks must lie:

\begin{enumerate}
\setlength{\itemsep}{0pt}%
\setlength{\parskip}{0pt}%
    \item The case of perfect thermal contact between the blocks, which essentially corresponds to that of a continuous dump;
    \item The case of null thermal contact, in which, apart from the air flowing around them, the three blocks are thermally isolated from one another. The analysis for this case was focused on the first block, due to it being the one in which the vast majority of the total energy is deposited.   
\end{enumerate}

\begin{figure*}[htbp]
\centering
\includegraphics[width=0.75\textwidth]{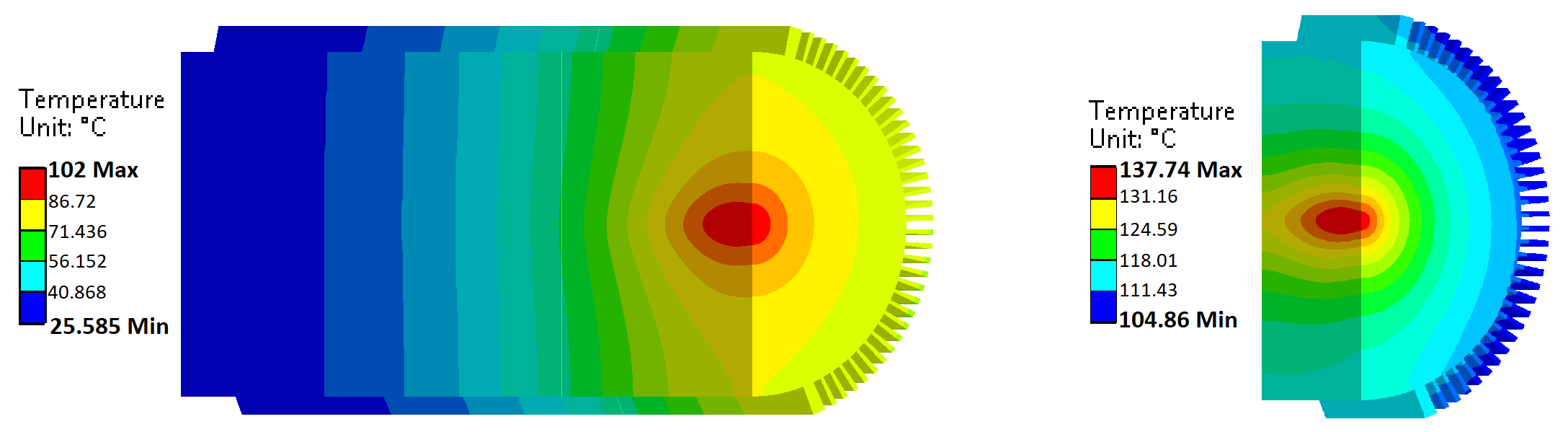}
\caption{Steady state temperature distributions achieved in the dump core for the two models taken into consideration under prolonged dumping of a 2~GeV NORMGPS beam with an intensity of $1\times 10^{14}$ppp at a 50\% dumping rate. On the left, the continuous dump model. On the right, the model assuming the first block isolated.}
\label{fig:2GeV10kW}
\end{figure*}

Similarly to the CFD analysis, a symmetric model was implemented in both cases.
The steady state temperature distributions computed for 2~GeV NORMGPS beam with an intensity of $1\times$10$^{14}$~ppp at a 50\% dumping rate are shown for the two considered models in Figure~\ref{fig:2GeV10kW}.  Steady state is achieved in around 1~h of this type of operation. Once this is achieved, a pulse impact causes a temporary increase in the maximum temperature of 10\degree~C. The maximum temperature that is reached in the cycle is, therefore, 148\degree~C, if the model with the first isolated block  is taken into consideration. 

\begin{figure*}[htbp]
\centering
\includegraphics[width=0.75\textwidth]{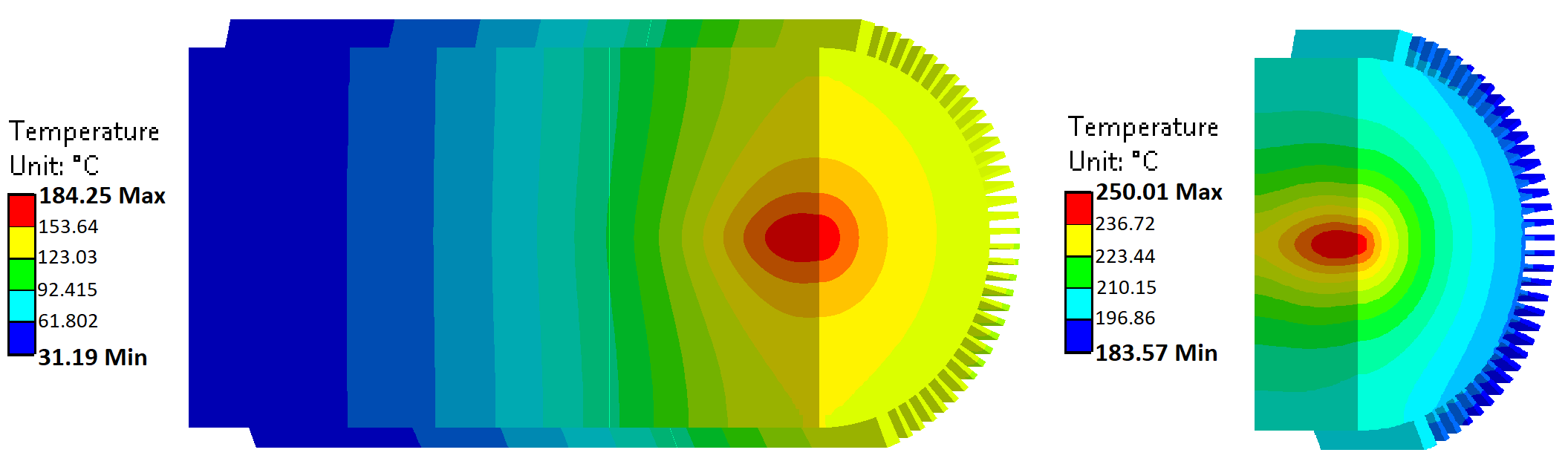}
\caption{Steady state temperature distributions achieved in the dump core for the two considered models under prolonged dumping of a 2~GeV NORMGPS beam with an intensity of $1\times$10$^{14}$~ppp at a 100\% dumping rate. On the left, the continuous dump model. On the right, the model with the first block isolated. Temperatures in $\degree$C.}
\label{fig:2GeV20kW}
\end{figure*}

It is also interesting to consider the temperature distribution that would be achieved in the dump core if the 2~GeV NORMGPS beam were to be dumped with a $1\times$10$^{14}$~ppp at a 100\% dumping rate. In these conditions, the power that is deposited on the dump is 19.34~kW, double that considered in the previous case. The steady state temperature profiles computed for these conditions are shown in Figure \ref{fig:2GeV20kW}.

Since the pulse intensity is the same as the previous case, each pulse would still cause a temporary increase of the maximum temperature in the dump core of 10$\degree$C. The maximum temperature in this case would then be 260$\degree$C if the model assuming the first isolated block is considered. As can be seen in Figure~\ref{fig:matpropCu},  the yield strength of the CuCr1Zr employed in the dump core decreases at this temperature only by around 15\% with respect to the yield strength at ambient temperature. This means therefore that, even considering a highly conservative thermal model such as the one with the first isolated block, impacted with a full intensity 2~GeV beam at a 100\% dumping rate, the maximum reached temperatures would be well within the maximum service temperatures of the dump core material.

\subsection{Mechanical response of the dump core to high intensity 2~GeV beams}
As previously detailed in Sec.~\ref{sec:beampar}, two among all the possible beams extracted from the PSB after LS2 were identified as the most critical for the operation of the dump core. The NORMGPS beam, considered in the previous section, has the smallest beam dimension among the beams that have an intensity of $1\times10^{14}$~ppp. The LHC25ns beam, while having a lower total intensity of $1.4\times10^{13}$~ppp, has a considerably smaller transversal beam size and, therefore, a higher maximum energy deposition density.
The model that was used to evaluate the dynamic mechanical response of the dump core to each of these beam impacts consists of a 2D axi-symmetric model of the first block of the dump core.



\begin{figure}[ht]
\centering
\includegraphics[width=0.45\textwidth]{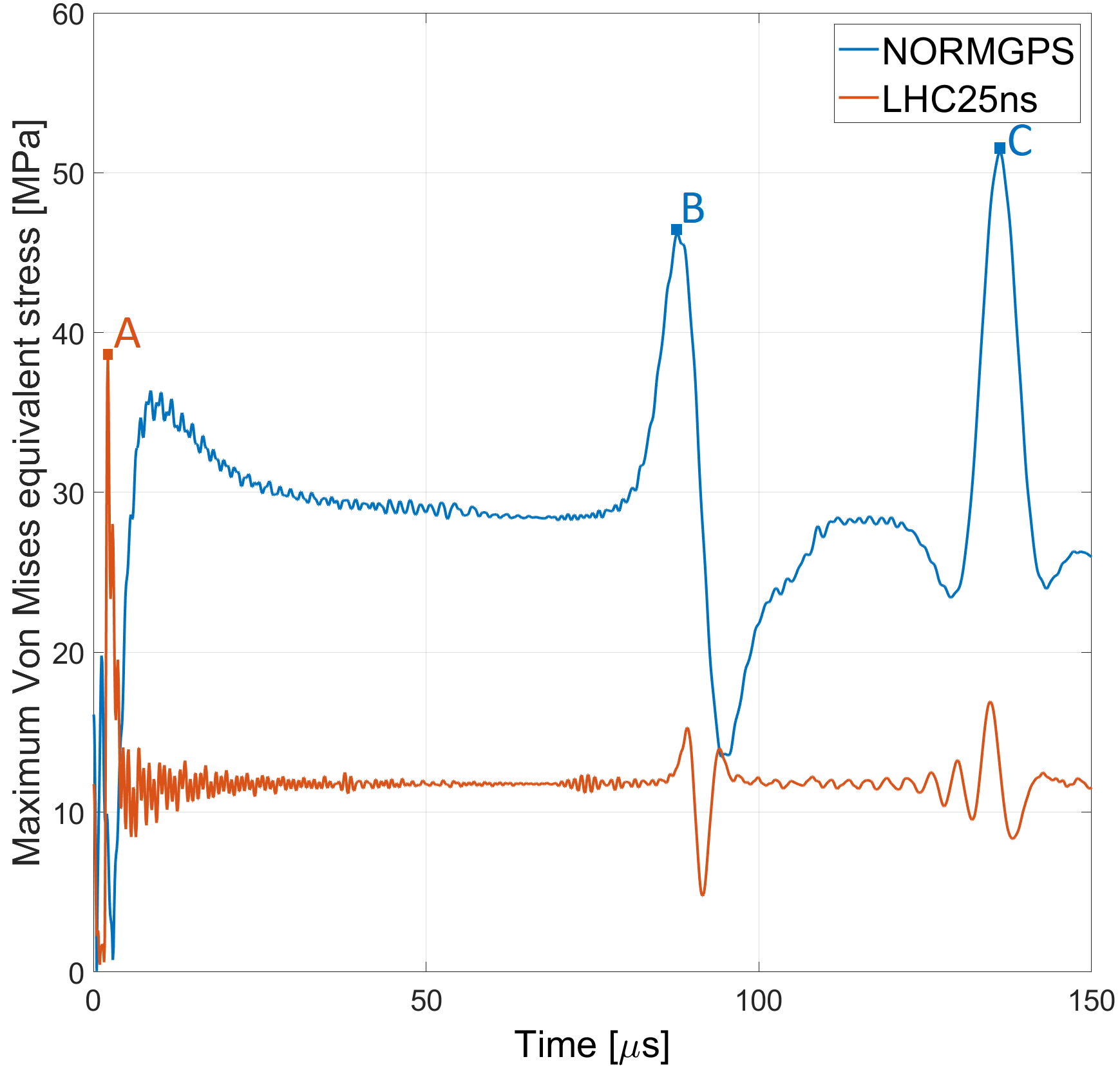}
\caption{Evolution over time of the maximum Von Mises stress in the dump core after impacts from the NORMGPS and LHC25ns beams.}
\label{fig:VMtime}
\end{figure}

\begin{figure}[ht]
\centering
\includegraphics[width=0.45\textwidth]{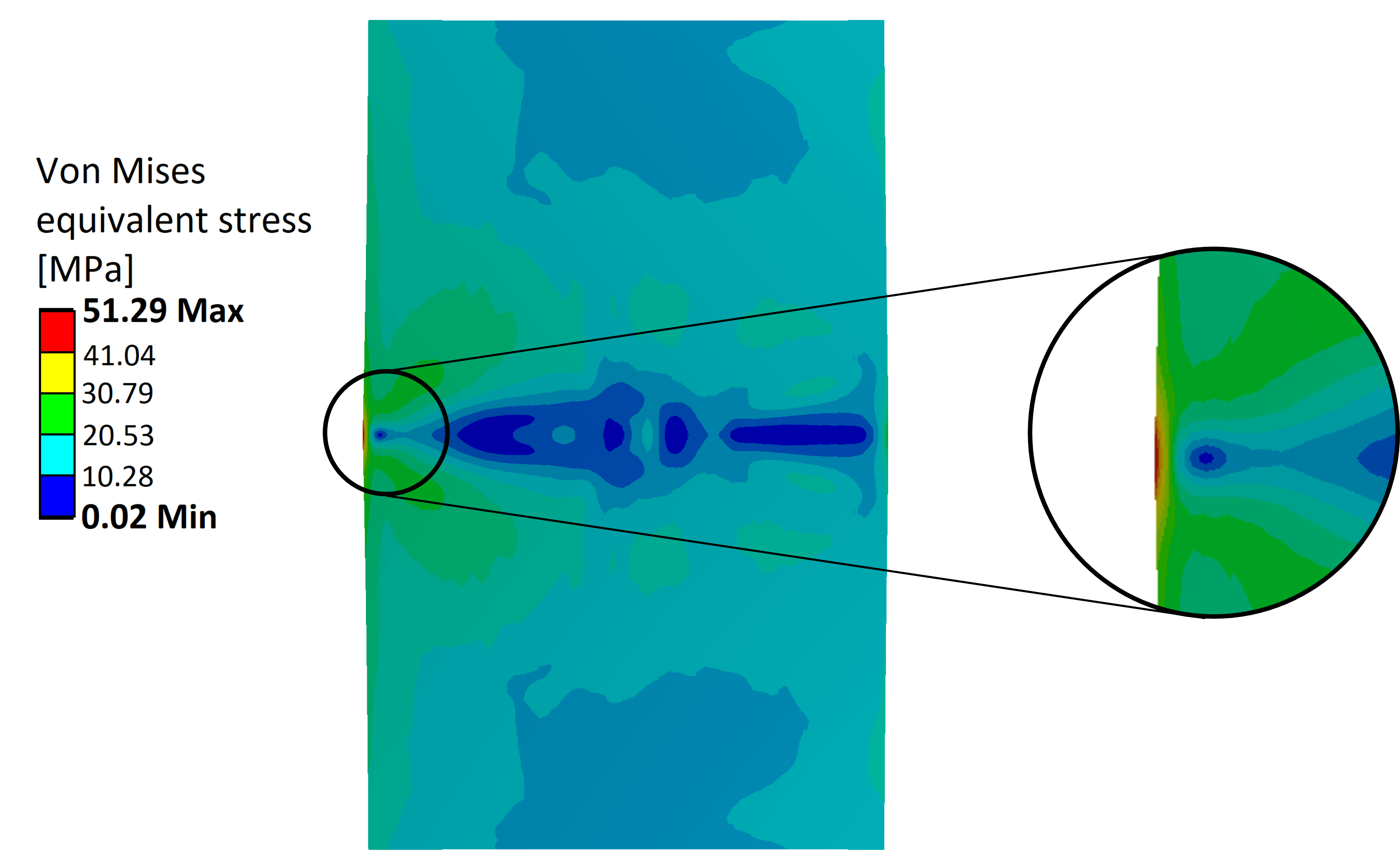}
\caption{Von Mises equivalent stress distribution achieved in the first block of the dump core after 136~$\mu$s from the initial NORMGPS beam impact (point C in the graphic in Figure~\ref{fig:VMtime}).}
\label{fig:VMNORMGPS}
\end{figure}

The time-dependent temperature response of the first dump core block to each beam impact, starting from a steady state operation achieved under NORMGPS beam dumping at a 50\% dumping rate, was first interpolated and applied to the mechanical model as a thermal load. The bunch structure of each beam was reproduced in order to more accurately model the dynamic response of the core to the beam impacts.
The evolution over time of the maximum Von Mises Equivalent Stress in the dump core caused by each beam impact is shown in Figure~\ref{fig:VMtime}. In both cases, the maximum stress takes place in the center of the upstream face of the block.
As can be seen, while the LHC25ns beam causes a higher peak Von Mises stress (point A in the graph, 39~MPa) just after the impact, the maximum values of stress are actually achieved some time after the NORMGPS impact. By studying the evolution over time of the pressure in the area surrounding the center of the upstream surface of the block, it was possible to identify the origin of these stress peaks. The second peak in the NORMGPS response (point B in the graph, 46.2~MPa at 87.8~$\mu$s) is caused by the radial stress wave that was originally generated on the axis of the dump core by the beam impact coming back towards the axis after having been reflected on the lateral surface. This can also be verified by noting that 87.8~$\mu$s also corresponds to the time that is necessary for the speed of sound in CuCr1Zr (4500~m/s) \cite{Meyers2007} to travel to the outer surface and back. The last peak (point C in the graph, 51.2~MPa at 136~$\mu$s) is instead induced by a Rayleigh surface wave, which travels more slowly than the speed of sound. The stress distribution at this time, corresponding to the case of the NORMGPS impact, is shown in Figure~\ref{fig:VMNORMGPS}. After having understood the causes of the higher values of the stress peaks in the NORMGPS impact, it is possible to state that these are  essentially due to the higher total amount of energy (nearly tenfold) that is deposited by this beam. As it was shown before, this higher amount of deposited energy causes, in turn, more intense stress waves to arise during the dynamic response of the dump core.

\begin{figure}[ht]
\centering
\includegraphics[width=0.45\textwidth]{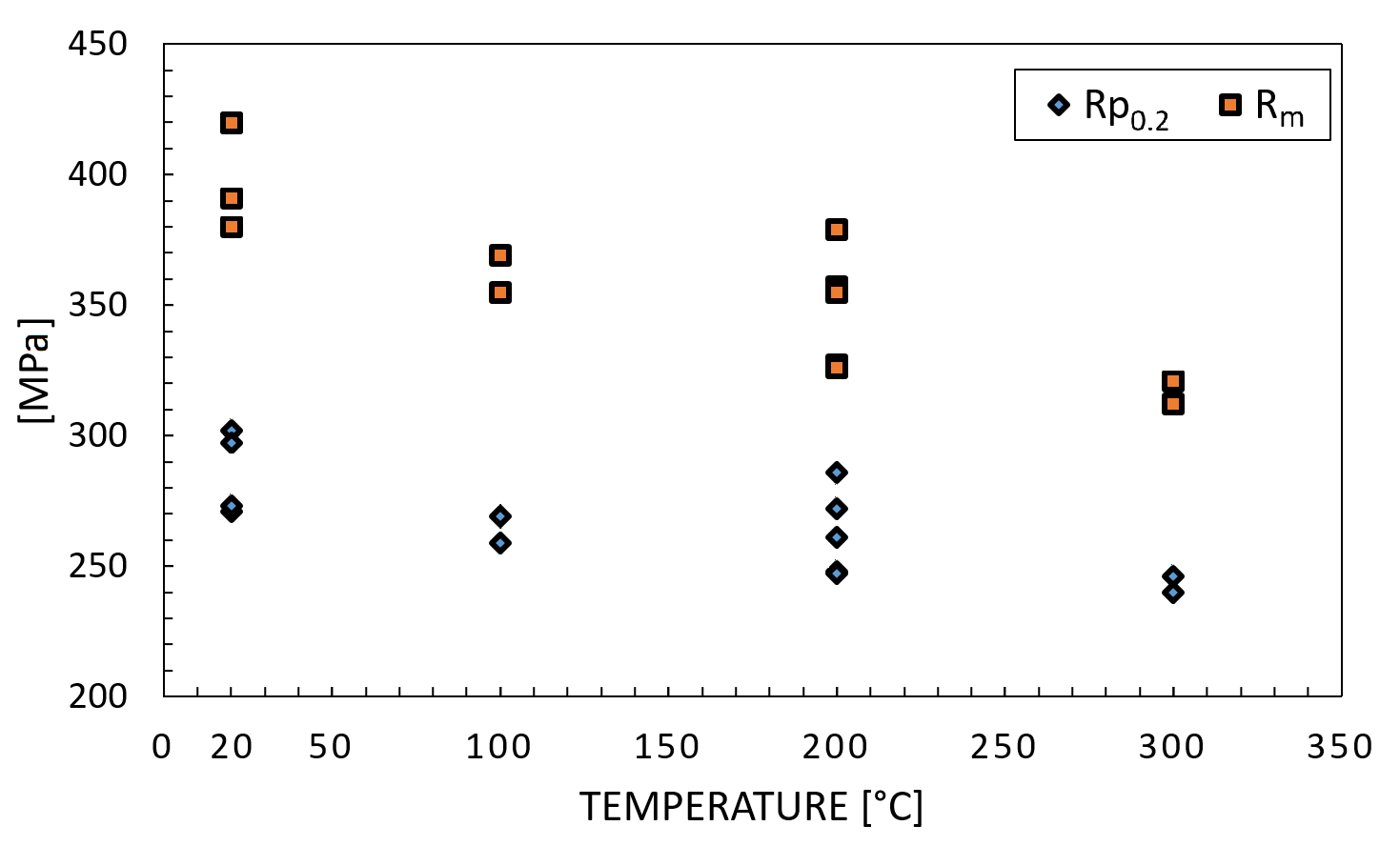}
\caption{Mechanical properties obtained from tensile tests on specimens that were extracted from notable position in a part that had the same dimension as the semi-finished product from which the dump core blocks were machined. ~\cite{Zollern2013}.}
\label{fig:matpropCu}
\end{figure}

It is finally important to compare these results with the mechanical properties of CuCr1Zr. Since this alloy is precipitation hardened, its final mechanical properties are achieved by solution annealing followed by cold working and ageing~\cite{li_physical_2012, barabash_specification_2011, deutsches_kupferinstitut_kupferdatenblatt_2005} and also depend on temper state and testing temperature.
Moreover, core hardenability also depends on the diameter of the semi-finished product from which the dump was machined. It was for this reason that a series of tensile tests were performed on specimens that were extracted from several positions in a part that had the same dimension as the semi-finished product from which the dump core blocks were machined. The results from these tests, for different testing temperatures, are summarized in Figure~\ref{fig:matpropCu}.

Even though the above comparison suggests that the design constraints could have been relaxed as the material could accept higher stresses and temperature, it should be noted that the dynamic loading conditions produced by the pulsed beam has an important effect on the fatigue life of the material. They were therefore maintained in order to achieve a higher safety factor on the fatigue behaviour of the core. Furthermore, the mechanical and thermal properties of the alloy can be degraded by the effect of a long exposure to irradiation and due to high integrated intensity to which the dump core will be subjected (estimated at 1.3$\times$10$\textsuperscript{21}$~protons during 25 years of operation of the dump).  The corresponding maximum value of the DPA (i.e. Displacements Per Atom) in the dump core, accumulated at the end of the 25 years of operation of the dump, was computed by means of FLUKA Monte Carlo simulations and corresponds to 0.5 DPA. Due to the substantial lack of data concerning the damage caused by proton irradiation on materials, it is difficult to evaluate the effect that such a high DPA value could have on the performance of the dump core.
Some first-order indications could be given by analyzing the results of neutron irradiation on the properties of the material.
In the study carried out in \cite{Kalinin2011}, CuCr1Zr fast-neutron irradiated up to 0.27 DPA exibited a limited increase in yield strength  and ultimate tensile strength due to radiation hardening of just \~6-10\%. The decrease in ductility was instead more pronounced.
As described in \cite{Eldrup1998},  electrical conductivity, which is directly proportional to the thermal diffusivity by means of the Wiedermann-Franz law, of CuCr1Zr irradiated up to 0.3 DPA in a fast-neutron reactor showed minor sensitivity to irradiation. Finally, the work carried out in \cite{Singh2005} shows how the irradiation of CuCr1Zr specimens up to 0.3 DPA determines an increase of the number of cycles to failure at a given stress amplitude. This corresponds to a fatigue life of 10\textsuperscript{4} cycles in a load-controlled creep fatigue test with a stress amplitude of 300~MPa, when tested at a temperature of 295~K and of approximately 250~MPa when tested at 573~K. This increase in fatigue life for a given stress amplitude further increases the safety factor on the fatigue life of the dump core.
The damage to the mechanical properties of CuCr1Zr caused by proton-beam irradiation will be further studied thanks to the post-irradiation examination of a capsule for material testing that was irradiated at Brookhaven National Laboratory's BLIP facility \cite{ammigan2018status}. The capsule, containing different layers of material specimens, among which CuCr1Zr, was subjected to an integrated intensity of $1.02\times10\textsuperscript{21}$  protons with an energy of 181~MeV. The maximum DPA value reached in the CuCr1Zr specimens as a result of this irradiation was 0.6 \cite{Espadanal2019}. The post-irradiation examination of the contents of the capsule for material testing is scheduled to take place during 2020.

\section{Replacement of the original dump}
\subsection{Radiation protection considerations}
As shown by the dose rate measurements that are summarized in Table~\ref{tab:RP},  the dose rate of the original dump core was particularly high after 40 years of irradiation. As a consequence of this, the dismantling procedure was defined following the ALARA (i.e. "As Low As Reasonably Achievable")~\cite{Forkel-Wirth:2013uza} safety principle. This approach aims at minimizing the dose of radiation to personnel by employing all reasonable methods.

For the case of the PSB dump, the application of the ALARA approach started with the creation of a map of the dose rate in the area where the work would be performed. On the basis of this, as well as of a detailed list of the actions to be taken and the time required for each of them, a work and dose planning document was compiled. This document contained a plan of the activities to be executed by each worker and the respective estimated absorbed dose. The workers also had to be trained for the tasks they performed and the dose received by each of them was continuously monitored. The collective dose accumulated by the personnel involved in the activity amounted to 1127 \textmu HSv.

\begin{table}[htbp]
\centering
\caption{Results of the dose rate measurements [\textmu Sv/h] performed on the main components of the original dump core-shielding assembly as they were being extracted from the cavity~\cite{Dumont2014}.}.
\label{tab:RP}
\begin{tabular}{lrr}
\hline \hline
Object                                         & Dose {[}\textmu Sv/h{]} & Distance {[}cm{]} \\ \hline
Dump core + beam pipe                          & 1100                                   & 160               \\
Beam pipe                                      & 35                                     & 160               \\
Shielding container of \\dump core and beam pipe & 750                                    & Contact           \\
Innermost shielding block                      & 2500                                   & 10                \\
Outermost shielding block                      & 30                                     & 10             \\  \hline
\end{tabular}
\end{table}

\subsection{Removal and dismantling of the original dump}
\label{sec:dismantling}


\subsubsection{Preparatory steps}
The removal and dismantling of the original dump first required three preparatory steps:
\begin{enumerate}
    \item Pre-shielding. In this phase, a steel cylinder, visible in Figure~\ref{fig:radiation_plug}, was inserted in the beam pipe in order to reduce the radiation from the dump core towards the outside of the dump cavity. In this way, the dose received by the workers during the following activities could be reduced from 120~$\mu$Sv/h down to 15~$\mu$Sv/h. This step was carried out in April 2013, four months prior to the following operations, and also proved useful for other activities in the area, prior to the removal of the dump;
    \item Temporary dismantling of the equipment in the BTM and BTY lines. In order to have the necessary space for extracting the original dump core-shielding assembly, the equipment in the BTM and BTY lines leading up to the dump cavity had to be temporarily dismantled and stored in a different facility;
    \item Lining of the area outside the dump with a plastic film. This coating, resistant to the mechanical stresses expected during the subsequent dismantling activities, was used to reduce the risk of contamination from the surface contamination present in the area.
\end{enumerate}

\subsubsection{Extraction and handling of the dump core and beam pipe}

\begin{figure}[ht]
%
\begin{subfigure}{0.35\linewidth}
\centering
\includegraphics[width=1\textwidth]{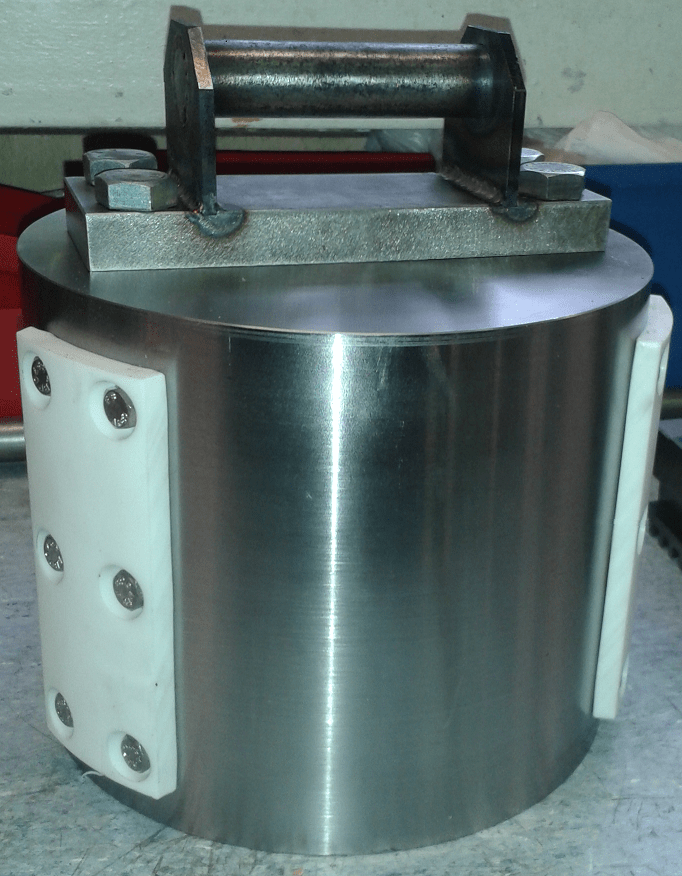}
\caption{}
\label{fig:radiation_plugOUT}
\end{subfigure}
\begin{subfigure}{0.9\linewidth}
\centering
\includegraphics[width=1\textwidth]{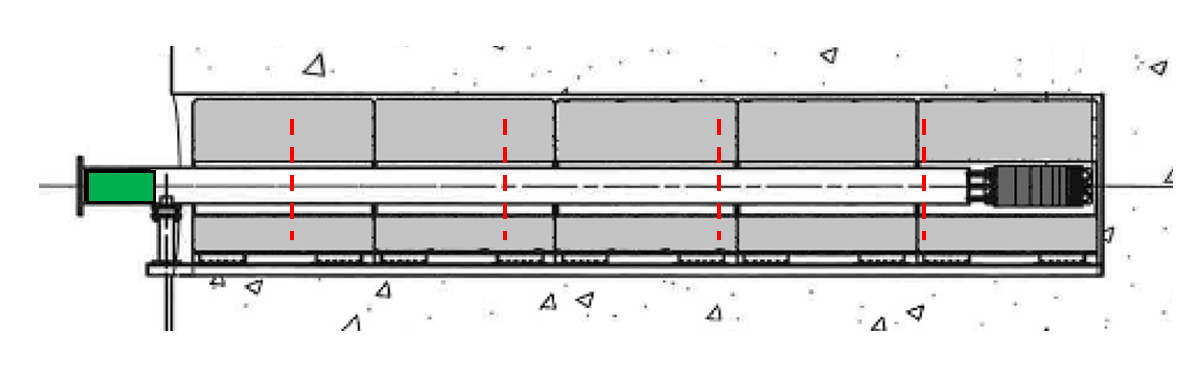}
\caption{}
\label{fig:radiation-plug-insertion}
\end{subfigure}
\caption{a) Stainless steel radiation shielding plug b) In green, insertion of the radiation plug at the end of the beam pipe of the original dump core-shielding assembly. The dashed lines indicate the position of the cuts performed by the remotely-operated saw on the beam pipe. }
\label{fig:radiation_plug}
\end{figure}

\begin{figure}[ht]
\centering
\includegraphics[width=0.4\textwidth]{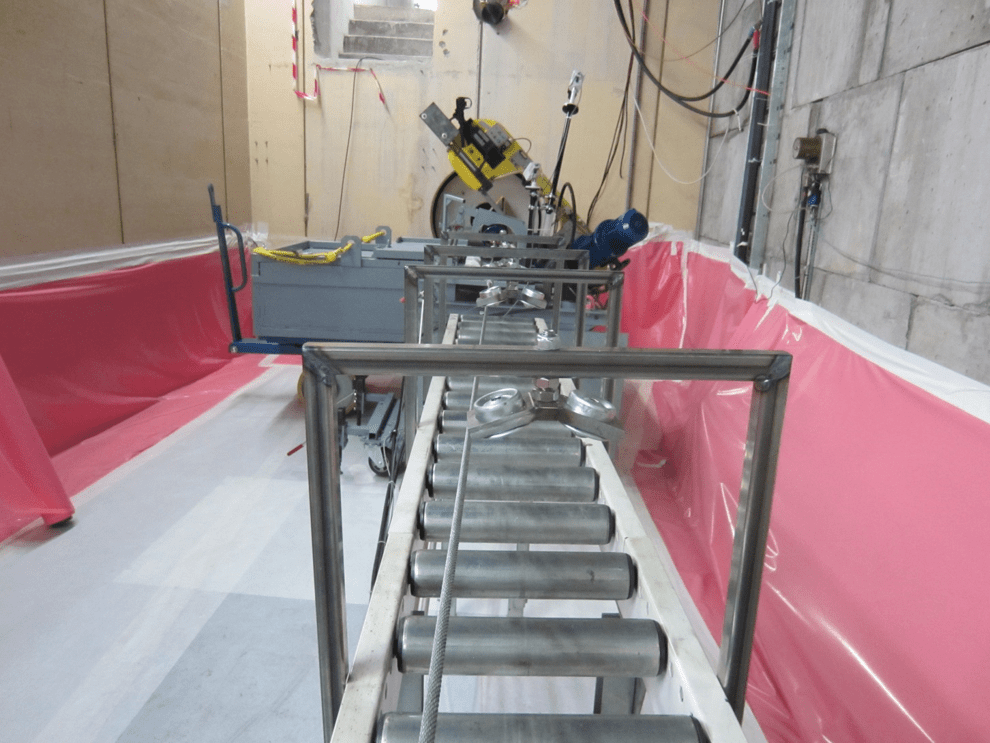}
\caption{Tooling for the extraction of the original dump core and beam pipe out of the dump cavity. In the foreground, the support structure with rollers and the winch cable. In the background, the remotely operated saw and the lead-shielded container.}
\label{fig:extraction-tooling}
\end{figure}

\begin{figure}[ht]
\centering
\includegraphics[width=0.4\textwidth]{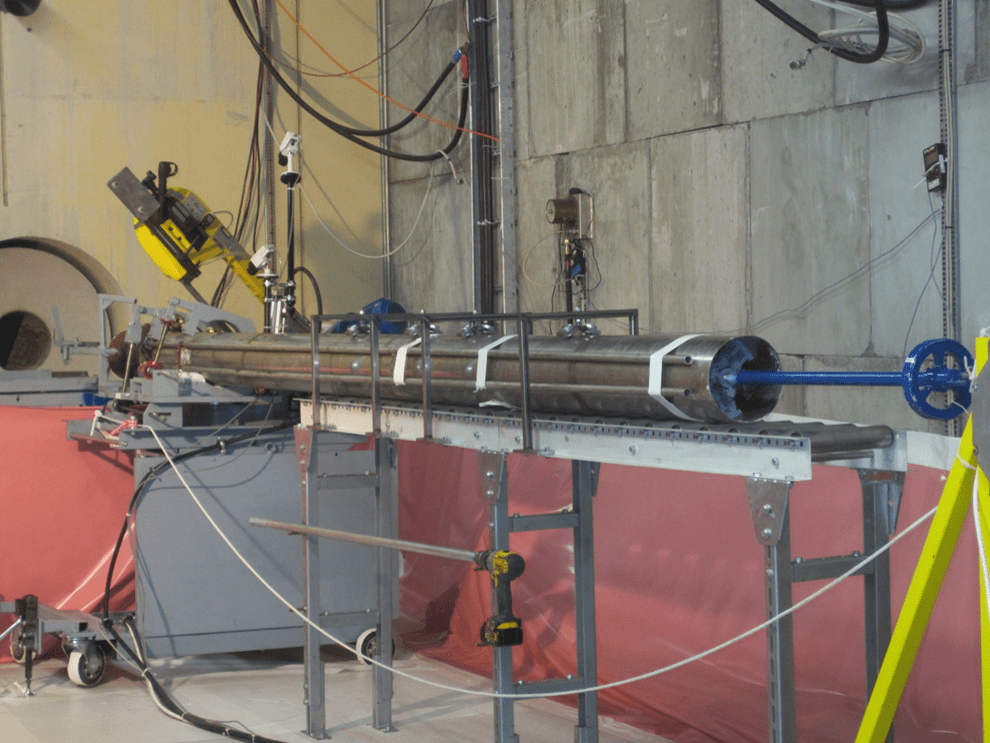}
\caption{Extraction of the original dump core and beam pipe from the dump cavity and on top of the structure with rollers. The blue clamping tool is visible on the right.}
\label{fig:old-dump-extraction}
\end{figure}

Once the preparatory steps were carried out, the tooling that was necessary for the extraction of the dump core was installed outside the cavity, as shown in Figure~\ref{fig:extraction-tooling}. 

The extremity of the beam pipe was attached to a winch by means of a clamping tool. Then, as shown in Figure~\ref{fig:old-dump-extraction}, by operating the winch, the dump core and the beam pipe were slid out of the cavity onto a support structure with rollers. After the full extraction of the assembly, a total of four cuts, indicated in Figure~\ref{fig:radiation_plug} with dashed lines, were performed on the beam pipe by means of a remotely-operated saw. After each cut, the assembly was pushed back towards the cavity, thereby making each cut part fall into a lead-shielded container, which is also visible in Figure~\ref{fig:extraction-tooling}.

After the cutting procedure was completed, the shielded container was transported to the radioactive waste storage facility at CERN for storage and the tooling for the extraction of the original dump core and beam pipe was disassembled.


\subsubsection{Extraction of the original shielding blocks}

\begin{figure}[ht]
\centering
\includegraphics[width=0.3\textwidth]{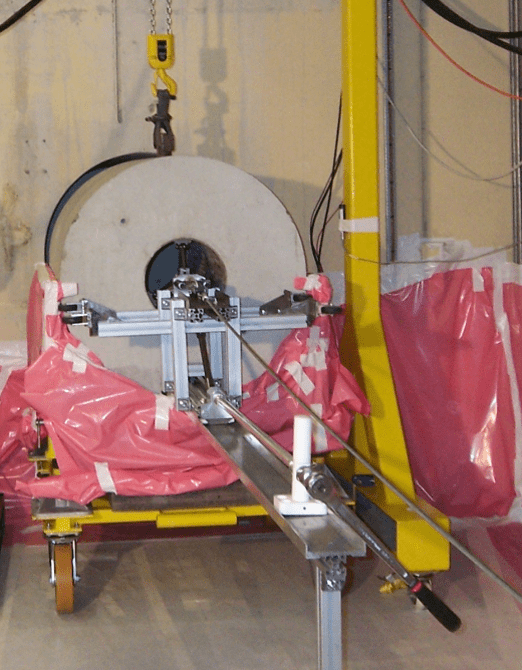}
\caption{Extraction of one of the original shielding blocks onto the support cradle.}
\label{fig:shielding-block-extraction}
\end{figure}

The five original shielding blocks were extracted by means of the tooling shown in Figure~\ref{fig:shielding-block-extraction}. As it can be seen, each shielding block was clamped from the inside by means of a clamping tool that was attached to the same winch cable that was used for the extraction of the dump core and beam pipe. Then, by operating the winch, the blocks were slid out of the cavity and onto a support cradle that was mounted outside of the cavity hole. This support cradle featured an extension of the steel rail on which the blocks were resting on the cavity (which can also be seen in Figure~\ref{fig:shielding-assembly}).

Once fully extracted, each block was loaded by means of a gantry crane inside individual shielded containers for transport and storage in a radioactive waste facility.

\subsection{Installation of the new dump core-shielding assembly}
\begin{figure}[ht]

\begin{subfigure}{0.495\linewidth}
\centering
\includegraphics[width=1\textwidth]{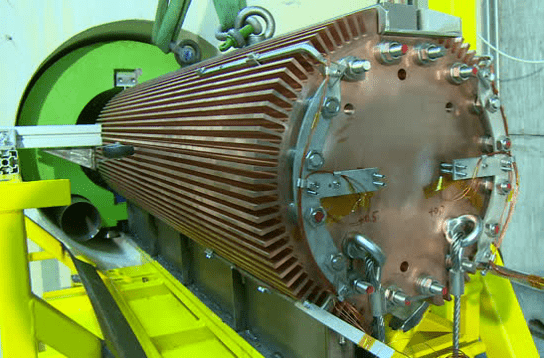}
\caption{}
\label{fig:dump_insertion}
\end{subfigure}
\begin{subfigure}{0.405\linewidth}
\centering
\includegraphics[width=1\textwidth]{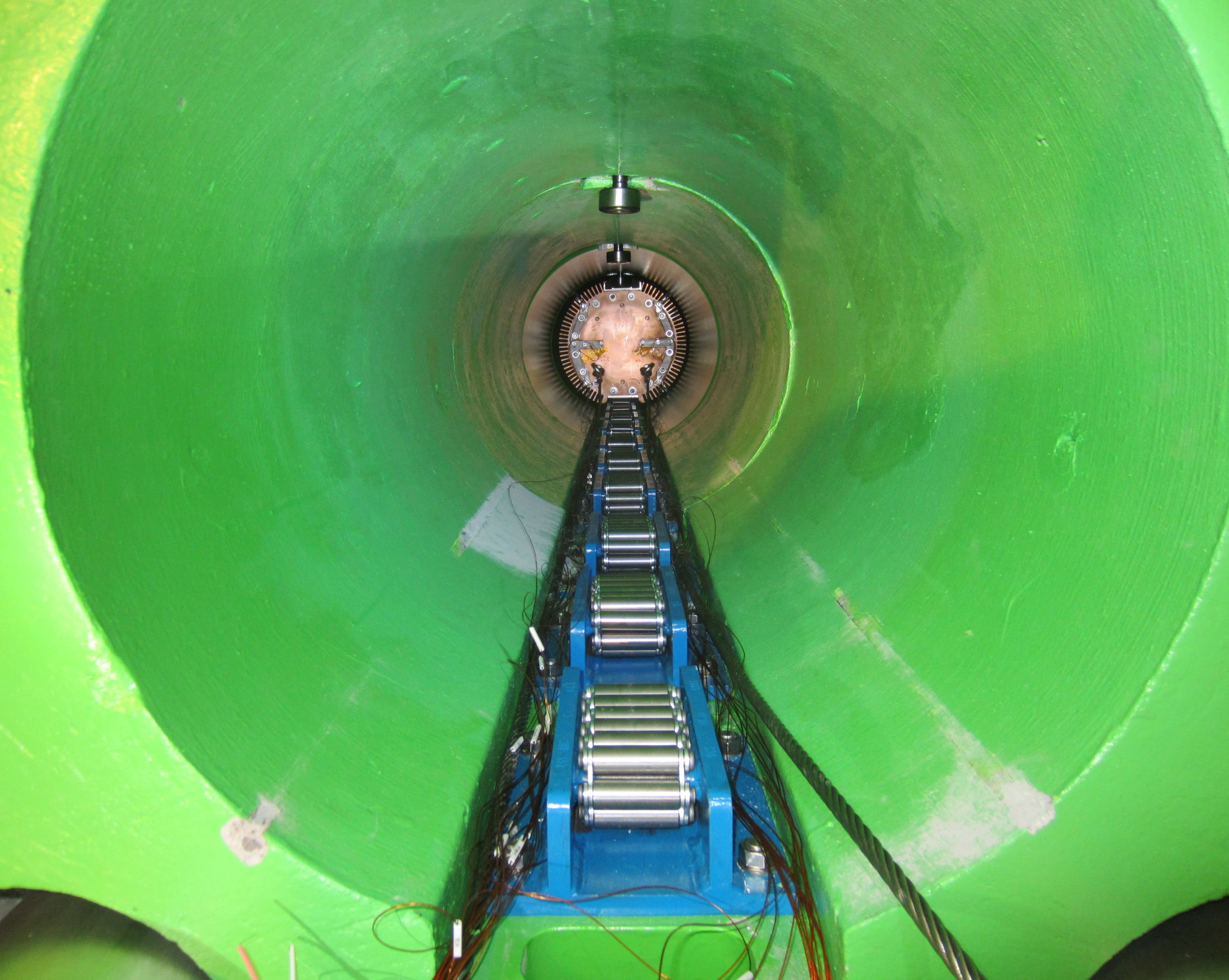}
\caption{}
\label{fig:Dump_inserted}
\end{subfigure}
\caption{a) New CuCr1Zr dump core prior to its insertion in the cavity.   b) The new CuCr1Zr dump core fully inserted in the dump cavity. Visible in blue, the rollers embedded in the shielding blocks that allowed the dump core to be slid to the end of the dump cavity.}
\label{fig:dump-insertion-double}
\end{figure}
\label{sec:installation}

Prior to the installation of the new dump core-shielding assembly in October 2013, the dump cavity was inspected and decontaminated. After this, samples of concrete were extracted from the lower part of the cavity. This, together with a structural analysis of the previously installed steel rail~\cite{garlasche_psb_2013}, was necessary to confirm that the dump cavity was indeed able to withstand the increased weight of the new dump core-shielding assembly, which is more than 50\% higher than before.
The installation then started with the pre-assembly of the air ducts for the ventilation and of the shielding blocks. This assembly was then progressively slid into the cavity by mounting each block on the same support cradle system that was previously used for the extraction of the old shielding. Finally, as shown in Figure~\ref{fig:dump-insertion-double}, the new dump core was slid to the end of the cavity by sliding it on the rollers that were embedded in the new shielding blocks.


\section{Monitoring and operational feedback}
\label{sec:operational-feedback}

The dump core and the respective cooling system were equipped with a set of sensors in order to assess their performance during their operation. These sensors include:

\begin{itemize}
    \setlength{\itemsep}{0pt}%
    \setlength{\parskip}{0pt}%
    \item A Pt100 probe (attached by thermal tab) and a type N thermocouple installed at each of twelve points on the dump core measure the temperature of the dump core. . Both types of sensors were polymide-insulated;
    \item Six Pt100 sensors measure the outlet temperature of the air;
    \item A calorimetric sensor gathers data on the flow and the temperature of the air before it enters the dump cavity;
    \item A Beam Current Transformer (BCT) is also installed upstream of the dump and measures the intensity of the dumped beam.
\end{itemize}

\begin{figure}[ht]
\centering
\includegraphics[width=0.45\textwidth]{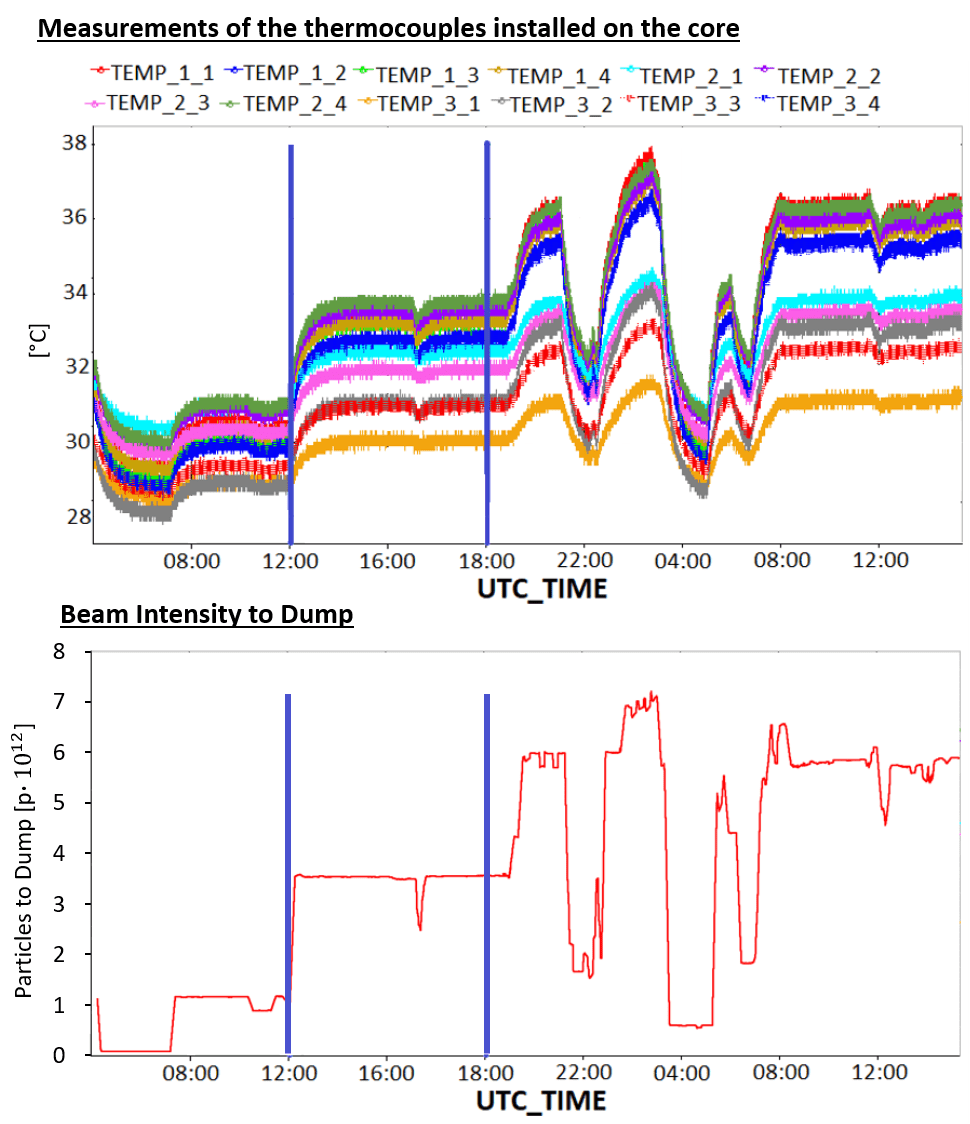}
\caption{Above: Measurements of the thermocouples  installed on the dump core  during the time period between 04:00 on the 11th of March, 2018 and 15:00 of the following day. Below: Ten-minute moving average of the intensity extracted to the dump in the corresponding time period.}
\label{fig:timber}
\end{figure}

The data measured by these sensors is collected in CERN's \textit{TIMBER} logging system~\cite{Billen2006}. 
An example of the data that is measured by the thermocouples installed on the dump core and of the average intensity of the beam extracted to the dump is shown in Figure \ref{fig:timber} for the time period from 04:00 on the 11th of March, 2018 to 15:00 of the following day. The measurements of the Pt100 sensors that were installed on the dump core, not displayed in the Figure, clearly showed signs of malfunction for most of these sensors.
Despite the fact that the highlighted period exhibited some of the highest intensities of the beam dumped  between LS1 and LS2, the  1.4~GeV beams were sent to the dump core with an average intensity of only $3.3\times10^{12}$~ppp. This is only a small fraction of the design intensity that was considered in Sec. \ref{sec:Thermal-analysis} for the 2~GeV NORMGPS beam.
The beam impacts during this period deposited an average power of 500~W over the 1.2~s-long cycle. This power, which corresponds to only 5\% of the 9.7~kW design power that was considered in Sec. \ref{sec:Thermal-analysis}, was evaluated by means of FLUKA Monte Carlo simulations considering a 1.4~GeV NORMGPS beam with the same characteristics as those that were measured during that week of operation. 

It is also important to mention that, due to a temporary malfunction of the cooling system, the inlet temperature of the air in this period was 29\degree~C. Considering the fact that the maximum temperature that was registered by the thermocouples during this period was 34\degree$C$, it can be stated that the increase in temperature due to the repeated beam impacts corresponded only to 5\degree$C$.
This temperature increase is of the same order of magnitude of the 2.5$\degree$C accuracy that can be expected for the type N thermocouples that are installed on the dump core.
Nevertheless, it was considered interesting to compare these measurements to the results given by the FEM model that was used to guide the design of the dump core (which was described in Sec. \ref{sec:Thermal-analysis}) for these lower-intensity beam impacts.
The thermocouples that are installed on the upstream face of the dump core were chosen for this comparison.

\begin{figure}[ht]
\centering
\includegraphics[width=0.3\textwidth]{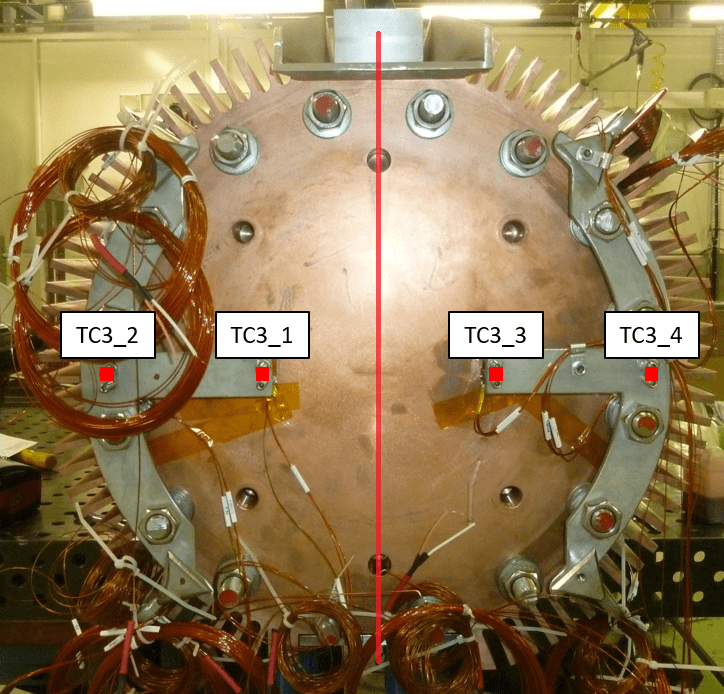}
\caption{Positions and labels of the four thermocouples that are placed on the upstream face of the dump core. The vertical red line corresponds to the projection of the symmetry plane that was implemented in the FEM simulations.}
\label{fig:Sensors}
\end{figure}

The position of these thermocouples, along with their respective labels, are shown in Figure~\ref{fig:Sensors}.

The temperatures that were registered by these thermocouples, along with the temperatures computed in the corresponding points by the thermal FEM simulation, are summarized in Table~\ref{table:Comparison}. 

\begin{table}[ht]
\centering
\caption{Average temperatures registered by the thermocouples placed on the upstream surface of the dump core in the time period from 12:00 to 20:00 of the 11th of March 2018. An accuracy of 2.5$\degree$C is expected for these sensors. Next to the measurements, the temperatures registered in the corresponding points of the two FEM models that were considered. Since the FEM models are symmetric with respect to the vertical plane, only one FEM result corresponds to each couple of thermocouple measurements.  Temperatures are reported in $\degree$C.}
\label{table:Comparison}
\begin{tabular}{lrr|r|r}
\hline \hline
 & \multicolumn{1}{l}{Thermocouples} & \multicolumn{1}{l}{} & \multicolumn{2}{|c}{FEM} \\\hline
ID & Temp.         & Std. Dev               & Continuous Dump    & I block isolated  \\\hline
TC3\_1          & 30.1                 & 0.3                & 32.8       & 34.8         \\
TC3\_3          & 31                   & 0.4                &            &              \\\hline
TC3\_2          & 31.2                 & 0.5                & 32.3       & 34.2         \\
TC3\_4          & 32.9                 & 0.6                &            &        \\\hline     
\end{tabular}%
\end{table}

As can be seen in the table, the temperatures calculated by the two FEM models that were considered in Sec.~\ref{sec:Thermal-analysis} are, within the aforementioned uncertainties, comparable to those that were registered by the thermocouples installed on the dump core.
After LS2, the higher temperatures achieved in the dump core as a consequence of the higher energy and intensities of the dumped beams will allow a more accurate comparison of the measured temperatures with the results of the FEM models.

\section{Conclusions}
In order to cope with the more intense and powerful beam expected after the PS Booster upgrade during LS2, thorough calculations were performed to produce a robust, conservative (hence reliable) beam dump design. In 2013 the new dump was installed in the same cavity as the original dump, following strict radio-protection protocols to minimize the dose to personnel. The new dump consists of three cylindrical blocks of CuCr1Zr, held together by screws and spring washers. The assembly is cooled by forced air convection, injected into the cavity where the dump is located and flushed back out into the accelerator tunnel. To maximize the cooling, fins were included in the design. The dump has operated between LS1 and LS2 well below its design parameters and is ready to handle the increased load expected after 2021. A testing run at full power is foreseen in late 2021 or early 2022.

\section{Acknowledgment}
The authors would like to acknowledge the LIU-PSB Project team at CERN, for its significant contribution to develop, support, review and improve this dump design. The authors are especially thankful to the following groups at CERN: EN-STI, EN-MME, EN-CV, EN-HE, BE-OP.

\bibliography{PSB-Dump}

\begin{thebibliography}{26}
\providecommand{\natexlab}[1]{#1}
\providecommand{\url}[1]{\texttt{#1}}
\expandafter\ifx\csname urlstyle\endcsname\relax
  \providecommand{\doi}[1]{doi: #1}\else
  \providecommand{\doi}{doi: \begingroup \urlstyle{rm}\Url}\fi

\bibitem[Baribaud and Metzger(1973)]{baribaud_800_1973}
G.~Baribaud and C.~Metzger.
\newblock The 800 {MeV} {Measurement} {Line} of the {CERN} {PS} {Booster}.
\newblock \emph{IEEE Transactions on Nuclear Science}, 20\penalty0
  (3):\penalty0 659--661, 1973.
\newblock ISSN 0018-9499.
\newblock \doi{10.1109/TNS.1973.4327209}.

\bibitem[Hanke(2013)]{hanke_past_2013}
K.~Hanke.
\newblock Past and present operation of the {CERN} {PS} {Booster}.
\newblock \emph{International Journal of Modern Physics A}, 28\penalty0
  (13):\penalty0 1330019--1--1330019--25, 2013.

\bibitem[Damerau et~al.(2014)Damerau, Funken, Garoby, Gilardoni, Goddard,
  Hanke, Lombardi, Manglunki, Meddahi, Mikulec, Rumolo, Shaposhnikova,
  Vretenar, and Coupard]{Damerau:1976692}
H~Damerau, A~Funken, R~Garoby, S~Gilardoni, B~Goddard, K~Hanke, A~Lombardi,
  D~Manglunki, M~Meddahi, B~Mikulec, G~Rumolo, E~Shaposhnikova, M~Vretenar, and
  J~Coupard.
\newblock {LHC Injectors Upgrade, Technical Design Report, Vol. I: Protons}.
\newblock Technical Report CERN-ACC-2014-0337, CERN, Dec 2014.
\newblock URL \url{https://cds.cern.ch/record/1976692}.

\bibitem[ISO 9330-3:1997()]{ISO9930}
ISO 9330-3:1997.
\newblock {Welded steel tubes for pressure purposes — Technical delivery
  conditions — Part 1: Unalloyed steel tubes with specified room temperature
  properties}.
\newblock Standard, International Organization for Standardization, Geneva, CH,
  March 1997.

\bibitem[Sarrio~Martinez(2013)]{sarrio_martinez_psb_nodate}
A.~Sarrio~Martinez.
\newblock {PSB} dump: {ALARA} procedure meeting.
\newblock Technical Report INDICO, LIU-PSB, CERN, Geneva, 2013.

\bibitem[Bartmann and Mikulec(2012)]{bartmann_ps_2012}
W.~Bartmann and B.~Mikulec.
\newblock {PS} {Booster} {Dump} {Upgrade}.
\newblock Technical Report PBU-T-ES-0002, EDMS 1229493, CERN, Geneva, 2012.

\bibitem[Lombardi(2019)]{Lombardi2019}
A~M Lombardi.
\newblock {Linac4 Commissioning}.
\newblock In \emph{29th Linear Accelerator Conf., LINAC2018}, number September,
  pages 658--662, Beijing, China, 2019.
\newblock ISBN 9783954501946.
\newblock \doi{10.18429/JACoW-LINAC2018-TH1P01}.

\bibitem[deu(2005)]{deutsches_kupferinstitut_kupferdatenblatt_2005}
Kupferdatenblatt {CuCr}1zr.
\newblock Technical report, Deutsches Kupferinstitut, 2005.

\bibitem[Garoby et~al.(2017)Garoby, Vergara, Danared, Alonso, Bargallo,
  Cheymol, Darve, Eshraqi, Hassanzadegan, Jansson, Kittelmann, Levinsen,
  Lindroos, Martins, Midttun, Miyamoto, Molloy, Phan, and et~al.]{Garoby_2017}
Roland Garoby, A~Vergara, H~Danared, I~Alonso, E~Bargallo, B~Cheymol, C~Darve,
  M~Eshraqi, H~Hassanzadegan, A~Jansson, I~Kittelmann, Y~Levinsen, M~Lindroos,
  C~Martins, {\O}~Midttun, R~Miyamoto, S~Molloy, D~Phan, and A~Ponton et~al.
\newblock The european spallation source design.
\newblock \emph{Physica Scripta}, 93\penalty0 (1):\penalty0 014001, dec 2017.
\newblock \doi{10.1088/1402-4896/aa9bff}.
\newblock URL \url{https://doi.org/10.1088%2F1402-4896%2Faa9bff}.

\bibitem[Lee et~al.(2017)Lee, Olsson, Eshraqi, Miyamoto, M{\"{o}}ller, Shea,
  Thomas, Wilborgsson, Nilsson, Molloy, Levinsen, and Sordo]{Lee2017}
Y~Lee, A~Olsson, M~Eshraqi, R~Miyamoto, M~M{\"{o}}ller, T~Shea, C~Thomas,
  M~Wilborgsson, S~Ghatnekar Nilsson, S~Molloy, Y~Levinsen, and F~Sordo.
\newblock {Working Concept of 12.5 kW Tuning Dump at ESS, paper THPVA065}.
\newblock \emph{Proceedings of IPAC2017}, pages 4591--4594, 2017.

\bibitem[Ferrari et~al.(2005)Ferrari, Sala, Fass{\`o}, and
  Ranft]{ferrari_fluka:_2005}
A.~Ferrari, P.R. Sala, A.~Fass{\`o}, and J.~Ranft.
\newblock {FLUKA}: a multi-particle transport code.
\newblock Technical Report CERN- 2005-10 INFN/TC 05/11, SLAC-R-773, CERN,
  Geneva, 2005.

\bibitem[Mason(2012)]{Mason2012}
G.~Mason.
\newblock {Functional Specification LIU-PSB Beam Dump Cooling System, EDMS
  1245969}.
\newblock Technical report, CERN, 2012.

\bibitem[Mason(2013)]{mason_liu-psb_2013}
G.~Mason.
\newblock {LIU}-{PSB} beam dump cooling system.
\newblock Technical Report EDMS 1250483, CERN, Geneva, 2013.

\bibitem[Meyers(2007)]{Meyers2007}
Marc~A Meyers.
\newblock \emph{{Dynamic Behavior of Materials}}.
\newblock Wiley, 2007.
\newblock ISBN 047158262X.

\bibitem[Li and Zinkle(2012)]{li_physical_2012}
Meimei Li and Steven~J. Zinkle.
\newblock Physical and {Mechanical} {Properties} of {Copper} and {Copper}
  {Alloys}.
\newblock \emph{Konings R.J.M., Comp. Nuc. Mat.}, 4:\penalty0 667--690, January
  2012.

\bibitem[Barabash et~al.(2011)Barabash, Kalinin, Fabritsiev, and
  Zinkle]{barabash_specification_2011}
Vladimir Barabash, G.~Kalinin, S.~Fabritsiev, and S.J. Zinkle.
\newblock Specification of {CuCrZr} alloy properties after various
  thermo-mechanical treatments and design allowables including neutron
  irradiation effects.
\newblock \emph{Journal of Nuclear Materials}, 417\penalty0 (904-907), 2011.

\bibitem[Zol(2013)]{Zollern2013}
{Certificate of mechanical properties of the copper for the PSB dump core, EDMS
  2257340}.
\newblock Technical report, CERN, 2013.

\bibitem[Kalinin et~al.(2011)Kalinin, Artyugin, Yvseev, Shushlebin, Sinelnikov,
  and Strebkov]{Kalinin2011}
G.~M. Kalinin, A.~S. Artyugin, M.~V. Yvseev, V.~V. Shushlebin, L.~P.
  Sinelnikov, and Yu~S. Strebkov.
\newblock {The effect of irradiation on tensile properties and fracture
  toughness of CuCrZr and CuCrNiSi alloys}.
\newblock In \emph{Journal of Nuclear Materials}, 2011.
\newblock \doi{10.1016/j.jnucmat.2011.02.036}.

\bibitem[Eldrup and Singh(1998)]{Eldrup1998}
M.~Eldrup and B.~N. Singh.
\newblock {Influence of composition, heat treatment and neutron irradiation on
  the electrical conductivity of copper alloys}.
\newblock \emph{Journal of Nuclear Materials}, 258-263\penalty0 (PART 1
  A):\penalty0 1022--1027, 1998.
\newblock ISSN 00223115.
\newblock \doi{10.1016/S0022-3115(98)00390-0}.

\bibitem[Singh et~al.(2005)Singh, Johansen, Li, and Stubbins]{Singh2005}
B~N Singh, B~S Johansen, M~Li, and J~F Stubbins.
\newblock \emph{{Creep-fatigue deformation behaviour of OFHC-copper and CuCrZr
  alloy with different heat treatments and with and without neutron
  irradiation}}, volume 1528.
\newblock 2005.
\newblock ISBN 87-550-3465-9.
\newblock URL
  \url{http://inis.iaea.org/search/search.aspx?orig{\_}q=RN:36109006}.

\bibitem[Ammigan and Hurh(2018)]{ammigan2018status}
K.~Ammigan and P.~Hurh.
\newblock Status and update of the radiate collaboration r\&d program, 2018.

\bibitem[Espadanal et~al.(2019)Espadanal, Vlachoudis, and
  Calviani]{Espadanal2019}
J.C. Espadanal, V.~Vlachoudis, and M.~Calviani.
\newblock {Studies of damage to materials for the BLIP tests and BDF target,
  using FLUKA Monte Carlo code. EDMS: 2114686}.
\newblock Technical report, CERN, 2019.

\bibitem[Forkel-Wirth et~al.(2013)Forkel-Wirth, Roesler, Silari,
  Streit-Bianchi, Theis, Vincke, and Vincke]{Forkel-Wirth:2013uza}
Doris Forkel-Wirth, Stefan Roesler, Marco Silari, Marilena Streit-Bianchi,
  Christian Theis, Heinz Vincke, and Helmut Vincke.
\newblock {Radiation protection at CERN}.
\newblock In \emph{{Proceedings, CERN Accelerator School on High Power Hadron
  Machines (CAS 2011): Bilbao, Spain, May 24-June 02, 2011}}, pages 415--436,
  2013.
\newblock \doi{10.5170/CERN-2013-001.415}.
\newblock [,415(2013)].

\bibitem[Dumont(2014)]{Dumont2014}
G.~Dumont.
\newblock {Changement du Dump du Booster, EDMS 1393879}.
\newblock Technical report, CERN, 2014.

\bibitem[Garlasche(2013)]{garlasche_psb_2013}
M.~Garlasche.
\newblock {PSB} dump insertion.
\newblock Technical Report EDMS 1302713, CERN, Geneva, 2013.

\bibitem[Billen and Roderick(2006)]{Billen2006}
R~Billen and C~Roderick.
\newblock {The LHC Logging Service Capturing , storing and using time-series
  data for the world ' s largest scientific instrument}.
\newblock Technical Report November, 2006.
\newblock URL
  \url{https://twiki.cern.ch/twiki/bin/view/ABPComputing/LhcDataStorage}.

\end{thebibliography}

\end{document}